\documentclass[
  journal=pasa,
  manuscript=research-paper, 
  year=2020,
  volume=37,
]{cup-journal}

\usepackage{microtype,siunitx,booktabs}
\usepackage{amsmath, multirow}
\usepackage[normalem]{ulem} 
\usepackage{soul,xcolor} 
\setstcolor{red} 
\def\HI{{\rm H\,{\textsc{\romannumeral 1}}}}
\def\HII{{\rm H\,{\textsc{\romannumeral 2}}}}
\def\degr{{$\hbox{$^\circ$}$}}
\def\arcmin{{$^\prime$}}
\def\arcsec{{$^{\prime\prime}$}}
\def\farcs{{$.\!\!^{\prime\prime}$}}
\def\fs{{$.\!\!^{\rm s}$}}
\def\fdg{{$.\!\!^\circ$}}

\title{The FAST Ultra-Deep Survey (FUDS): observational strategy, calibration and data reduction}

\author{Hongwei Xi}
\affiliation{CAS Key Laboratory of FAST, National Astronomical Observatories, Chinese Academy of Sciences, Datun Rd., Chaoyang District, Beijing 100101, China}
\alsoaffiliation{International Centre for Radio Astronomy Research (ICRAR), University of Western Australia, 35 Stirling Hwy, Crawley, WA 6009, Australia}
\alsoaffiliation{University of Chinese Academy of Sciences, Beijing 100049, China}
\alsoaffiliation{ARC Centre of Excellence for All Sky Astrophysics in 3 Dimensions (ASTRO 3D), Australia}

\author{Bo Peng}
\affiliation{CAS Key Laboratory of FAST, National Astronomical Observatories, Chinese Academy of Sciences, Datun Rd., Chaoyang District, Beijing 100101, China}
\email[Bo Peng]{pb@nao.cas.cn}

\author{Lister Staveley-Smith}
\affiliation{International Centre for Radio Astronomy Research (ICRAR), University of Western Australia, 35 Stirling Hwy, Crawley, WA 6009, Australia}
\alsoaffiliation{ARC Centre of Excellence for All Sky Astrophysics in 3 Dimensions (ASTRO 3D), Australia}

\author{Bi-Qing For}
\affiliation{International Centre for Radio Astronomy Research (ICRAR), University of Western Australia, 35 Stirling Hwy, Crawley, WA 6009, Australia}
\alsoaffiliation{ARC Centre of Excellence for All Sky Astrophysics in 3 Dimensions (ASTRO 3D), Australia}

\author{Bin Liu}
\affiliation{CAS Key Laboratory of FAST, National Astronomical Observatories, Chinese Academy of Sciences, Datun Rd., Chaoyang District, Beijing 100101, China}

\doi{10.1017/pasa.2020.32}

\received {dd Mmm YYYY}
\revised  {dd Mmm YYYY}
\accepted {dd Mmm YYYY}
\published{22 September 2020}


\begin{document}

\begin{abstract}

    
    The FAST Ultra-Deep Survey (FUDS) is a blind survey that aims for the direct detection of \HI\ in galaxies at redshifts $z<0.42$. The survey uses the multibeam receiver on the Five Hundred Meter Aperture Spherical Telescope (FAST) to map six regions, each of size 0.72 deg$^2$ at high sensitivity ($\sim 50 \mu$Jy) and high frequency resolution (23 kHz). The survey will enable studies of the evolution of galaxies and their \HI\ content with an eventual sample size of $\sim 1000$. We present the science goals, observing strategy, the effects of radio frequency interference (RFI)  at the FAST site, our mitigation strategies and the methods for calibration, data reduction and imaging as applied to initial data. The observations and reductions for the first field, FUDS0, are completed, with around 128 \HI\ galaxies detected in a preliminary analysis. Example spectra are given in this paper, including a comparison with data from the overlapping GAL2577 field of Arecibo Ultra-Deep Survey (AUDS).

\end{abstract}

\section{Introduction}\label{Sct_01}
    
    Hydrogen is the most abundant element in the Universe. In galaxies, hydrogen is found in three primary forms: neutral atomic (\HI), neutral molecular (H$_2$), and ionized (\HII). \HI\ is the most common phase of hydrogen found within late-type galaxies (disks or irregulars). \HI\ observations are also sensitive to low surface brightness, but gas-rich and dark-matter-rich galaxies, which are useful for studying dark matter concentration and constraining models of galaxy feedback and evolution. \HI\ observations, if extended over a significant redshift range, can provide valuable information on how the environmental and intrinsic factors affect galaxy formation and evolution change over cosmic time.
    
    Surveys with large sky coverage will detect more galaxies in a given total survey time. Large numbers of detections have therefore been made by the well-known shallow \HI\ surveys, namely the southern \HI\ Parkes All Sky Survey (HIPASS; \citealp{2001MNRAS.322..486B, 2004MNRAS.350.1195M}) and the northern Arecibo Legacy Fast ALFA survey (ALFALFA; \citealp{2005AJ....130.2598G}). HIPASS yielded 4,315 \HI\ galaxies over 21,341 deg$^2$ of southern sky by utilising a 13 beam receiver on the Parkes telescope. ALFALFA employed the 7-beam ALFA receiver on the Arecibo telescope to detect nearly 31,500 \HI\ galaxies over 7,000 deg$^2$ of northern sky. Whilst both surveys provided unprecedentedly large samples for statistical studies of HI in galaxies \citep{2005MNRAS.359L..30Z, 2016MNRAS.457.4393J, 2018MNRAS.477....2J}, they could not provide evolutionary information due to their limited depth (HIPASS, $z<0.04$; ALFALFA, $z<0.06$). New \HI\ surveys such as the Widefield ASKAP L-band Legacy All-sky Blind surveY (WALLABY; \citealp{2020Ap&SS.365..118K}) which will survey the southern sky, and the Commensal Radio Astronomy FAST survey (CRAFTS, \citealp{2018IMMag..19..112L, 2019SCPMA..6259506Z}) which will survey the northern sky, will continue this legacy and dramatically increase the sample sizes in the local Universe (both surveys will, in practice, be sensitivity-limited to $z<0.1$).
    
    Medium-deep \HI\ surveys such as the ASKAP Deep Investigation of Neutral Gas Origins (DINGO; \citealp{2009PRA...........M}) and the MeerKAT International GHz Tiered Extragalactic Exploration (MIGHTEE-HI, \citealp{2016mks..confE...6J}), have been proposed in order to target areas of 10's to 100's of square degrees out to deeper redshifts. However, in order to truly explore the \HI\ evolution of galaxies, smaller fields and higher sensitivity are required. Targeted or pencil beam surveys have been the preferred solution given limited observation time. Recent examples include the HIGHz project which employed the Arecibo telescope to search for \HI\ in 39 late-type galaxies at $z \sim 0.2$ \citep{2015MNRAS.446.3526C}. They did not find significant evolutionary trends between high-$z$ late-type galaxies and nearby galaxies. The Arecibo Ultra-Deep Survey (AUDS;  \citealp{2011ApJ...727...40F, 2015MNRAS.452.3726H}) was carried out to blindly search for \HI\ galaxies at $z<0.16$ in a limited 1.35 deg$^2$ area. Based on the final AUDS100 analysis, \citet{2021MNRAS.501.4550X} found evidence for the characteristic \HI\ mass (the `knee' in the \HI\ mass function) increasing with redshift. The ongoing COSMOS \HI\ large extragalactic survey (CHILES; \citealp{2016ApJ...824L...1F}) employs Very Large Array (VLA) to map $\sim 0.5$ deg$^2$ in the COSMOS field up to a higher redshift limit, $z=0.45$. In the near future, the Looking At the Distant Universe with the MeerKAT Array (LADUMA, \citealp{2016mks..confE...4B, 2018AAS...23123107B}), will map $\sim 2$ deg$^2$ with unprecedented sensitivity and larger redshift coverage ($z<1.4$). 

    The world's largest single-dish telescope, the Five-hundred-meter  Aperture  Spherical  Telescope (FAST, \citealp{2011IJMPD..20..989N}), having been operational for astronomical observations since 2019, is also available to explore the distant Universe with high sensitivity. Although its individual beam size is small, it is equipped with a 19-beam receiver at $L$-band, which dramatically increases its ability to survey the sky in \HI. Moreover, unlike the Parkes and Arecibo multibeam receivers, the FAST multibeam receiver can extend up to a redshift of $z \sim 0.4$. Therefore, a key scientific goal of FAST is to search for \HI\ in galaxies up to this redshift. In this paper, we introduce the FAST Ultra-Deep survey (FUDS), which will map six 0.72 deg$^2$ regions at high sensitivity at Declinations around +25\degr. The approximate on-source observing time is about 100 hr for each target area, which corresponds to a sensitivity of $\sim 50 \mu$Jy. We  estimate that FUDS will directly detect $\sim 1,000$ galaxies, and enable the studies of massive galaxies at high redshift as well as faint objects around Milky Way. The volume is sufficient to robustly study the evolution of the \HI\ mass function and the cosmic \HI\ density, $\Omega_{\HI}$.
    
    \begin{table}
        \centering
        \caption{Positions of the FUDS target fields. Observations for FUDS0 are complete and observation for FUDS1 have commenced. Observations for FUDS2 -- FUDS5 have not commenced. Their positions are preliminary.
        }
        \label{Tab_01}
        \begin{tabular}{lrrc}
            \hline
            \hline
            ID    & R.A.(J2000)  & Dec (J2000) & Area [deg$^2$] \\
            \hline
            FUDS0 & 08$^{\rm h}$17$^{\rm m}$12\fs0 & +22\degr10\arcmin48\arcsec & 0.72   \\
            FUDS1 & 15$^{\rm h}$40$^{\rm m}$52\fs2 & +18\degr42\arcmin00\arcsec & 0.72   \\
            FUDS2 & 23$^{\rm h}$14$^{\rm m}$55\fs2 & +27\degr06\arcmin00\arcsec & 0.72   \\ 
            FUDS3 & 01$^{\rm h}$03$^{\rm m}$00\fs0 & +21\degr33\arcmin00\arcsec & 0.72   \\
            FUDS4 & 10$^{\rm h}$51$^{\rm m}$19\fs2 & +24\degr52\arcmin48\arcsec & 0.72   \\
            FUDS5 & 13$^{\rm h}$40$^{\rm m}$09\fs6 & +20\degr37\arcmin48\arcsec & 0.72  \\
            \hline
            \hline
        \end{tabular}
    \end{table}
    
    In this paper, we introduce the scientific goals of FUDS in Section \ref{Sct_02}. The FAST telescope, region selection and observation strategy are described in Section \ref{Sct_03}. We present a study of the RFI at the FAST site in Section \ref{Sct_04} and the calibration method in Section \ref{Sct_05}. Using our pilot observations of the first two surveys fields (FUDS0 and FUDS1), data reduction of FUDS is introduced in Section \ref{Sct_06}. The impact of confusion is discussed in Section \ref{Sct_07}. We present the preliminary results from the first target field, FUDS0, in Section \ref{Sct_08}. Finally, a summary is given in Section \ref{Sct_09}.

\section{Scientific goals}\label{Sct_02}

    FUDS  aims at the direct detection of `high redshift' galaxies in \HI\ in a blind survey, and faint low-surface brightness features around nearby galaxies and filaments. The target sensitivity is 50~$\mu$Jy. Based on expectations from the \HI\ mass function, and initial results, FUDS is expected to find around 1,000 \HI\ galaxies and will address the following science questions.
    
    \subsection{The Cosmic Gas Density}\label{Sct_02_01} 
        
        The cosmic gas density, $\Omega_{\HI}(z)$ is a measure of how the global supply of cold gas available for star formation changes with redshift or cosmic time. Whilst the cosmic star formation rate (SFR) is known to have decreased by an order of magnitude from $z \sim 1$ to the present time \citep{1996ApJ...460L...1L, 1998ApJ...498..106M}, somewhat surprisingly, $\Omega_{\HI}(z)$ seems more constant. Recent semi-analytic \citep{2015MNRAS.453.2315K} and hydrodynamic \citep{2017MNRAS.467..115D} models have now been adjusted to reflect this slower evolution, but there remain modest differences between models at $z<1$ and enormous divergence at $z>1$ (see Figure 19 in \citealp{2021MNRAS.501.4550X}). Moreover, the observational and systematic errors in measurements of $\Omega_{\HI}(z>0.1)$ remain stubbornly significant. The main methods available to measure cosmic \HI\ density are:  (1) direct \HI\ spectral-line measurements of individual galaxies in a blind survey; (2) spectral stacking using optical redshifts; (3) intensity mapping; and (4) damped Lyman-$\alpha$ (DLA), or proxy measurements. Of these, the first is strongly preferred but, owing to the weakness of the 21-cm line, it has been difficult to accomplish at $z>0.1$. Most of $\Omega_{\HI}$ values at $z > 0.1$ (which are much less accurate than for $z<0.1$) use stacking \citep{2007MNRAS.376.1357L, 2013MNRAS.433.1398D, 2018MNRAS.473.1879R}, intensity mapping \citep{2010Natur.466..463C, 2013ApJ...763L..20M} or damped Lyman-$\alpha$ (DLA) \citep{2000ApJS..130....1R, 2009ApJ...696.1543P, 2020MNRAS.498..883G} methods. With the FUDS, we plan to perform the most accurate measurements to date for $\Omega_{\HI}$ at $z > 0.1$, without reference to optical surveys. Stacking methods will allow accurate measurements in the range $z=0.2-0.4$, but with a strong dependence on the selection and assumed evolutionary parameters of the corresponding optical input catalogue. 
        
    \subsection{The evolution of galaxies and their \HI\ content}\label{Sct_02_02}
        
        The study of the cold gas (and stellar) content of individual galaxies, and the effect of environment on gas supply and star-formation rate, is a key to better understand galaxy evolution. Again, due to the difficulties of individual galaxy detection, this has been challenging. \citet{2015MNRAS.446.3526C} measured the \HI\ content in 39 massive late type galaxies at $0.17 \leq z \leq 0.25$. Compared with scaling relations of galaxies at $z \sim 0$, they found that these galaxies are the analogues of extreme objects in the local Universe. As above, stacking techniques
        \citep{2001A&A...372..768C, 2001Sci...293.1800Z}
        can shed more light on the mean properties of galaxies over a larger redshift range. For example, \citet{2018MNRAS.473.1879R} measured a higher \HI\ mass-to-luminosity ratio for distant late-type galaxies as compared to samples at lower redshift. However, the requirement to use optical/IR samples adds extra measurement uncertainties. More importantly, it adds selection effects, such as imaging and spectroscopic incompleteness \citep{2006MNRAS.369...25J}, and systematic uncertainties due to redshift-dependent luminosity evolution \citep{2014MNRAS.445.2125M} and $k$-corrections \citep{2016OJAp....1E...3L}, and possible environmental dependence \citep{2014MNRAS.445.2125M}. There is also a dependence on optical luminosity and/or luminosity density functions which may have been derived from different optical samples. Perhaps the most important of these additional uncertainties is the tendency for optical catalogues to miss the substantial population of \HI-rich but optically low-surface brightness galaxies. This tendency is strongly reflected in the steeper low-mass slope in the \HI\ mass function compared with the optical luminosity or stellar mass functions \citep{2018A&A...619A..89B}.
        
        FUDS is intended to be deep enough to detect individual galaxies right up to the survey limit of $z \sim 0.4$, and in sufficient statistical numbers to allow an accurate study of the evolution of galaxies and their gas content. However, given the large physical size of the FAST beam at higher redshifts, confusion is a possible problem (see e.g. \citealp{2016MNRAS.455.1574J}). We will briefly discuss the impact of confusion on our results in Section \ref{Sct_07}.
        
    \subsection{The evolution of the \HI\ Mass and Velocity Functions}\label{Sct_02_03}
        
        As statistical quantities which relate to the number density of galaxies of different gas and halo mass, the HI Mass Function (HIMF) and velocity width function (HIWF) are important constraints for cosmological models. Thanks to the large sample sizes now available, the form of the HIMF and HIWF and their Schechter parameterisation \citep{1976ApJ...203..297S} have been greatly refined \citep{2018MNRAS.477....2J,2022MNRAS.509.3268O}.
        However, the HIMF beyond the local Universe is poorly known. Using the Arecibo Ultra-Deep Survey (AUDS), \citet{2015MNRAS.452.3726H} did not find any evolutionary trend in the preliminary AUDS60 sample. However for the full sample, \citet{2021MNRAS.501.4550X} found weak evidence for a `knee' mass which increases with redshift. This conclusion was drawn based on a sample of only 247 galaxies. With its larger sample size, and wider redshift range, FUDS will put more accurate constraints on evolutionary trends.
        
    \subsection{Environment dependence} 
        
        Many studies have explored the environmental dependence of the HIMF. For example, does the density ratio of low-mass and high-mass galaxies increase or decrease with local density? No common trends have yet been established, either for the slope of the low-mass end of the HIMF ($\alpha$) or the value of the characteristic mass ($M^*$). \citet{2002ApJ...567..247R} found a flatter $\alpha$ in an intermediate density region around the Virgo cluster compared with the field.\citet{2021MNRAS.501.2608B} also found a flat slope, $\alpha=-0.92$, in dense regions using \HI\ detections in the Ursa Major cluster. In contrast, \citet{2011ApJS..197...28P} found flatter $\alpha$ as the density decrease based on six loose groups. \citet{2005MNRAS.359L..30Z} found steeper $\alpha$ values in dense regions based on the full HIPASS sample. None of these studies found any change in $M^*$. \citet{2005ApJ...621..215S} used $\sim$ 3000 AGC galaxies to derive the HIMF and found constant $\alpha$ and decreasing $M^*$ in dense regions. Utilising the 40\% and 70\% ALFALFA samples, \citet{2014MNRAS.444.3559M} and \citet{2016MNRAS.457.4393J} discovered positive correlation between $M^*$ and density, respectively. But, neither found any obvious change in $\alpha$. \citet{2018MNRAS.477....2J} derived a steeper $\alpha$ and higher $M^*$ for non-cluster high-density regions from the 100\% ALFALFA sample. \citet{2019MNRAS.486.1796S} found flatter $\alpha$ in dense regions within the Parkes \HI\ Zone of Avoidance survey (HIZOA, \citealp{2005AJ....129..220D, 2016AJ....151...52S}). However, they did not find any change in $M^*$. By examing HI detections, and the form of the HIMF in different target regions, \HI\ surveys such as FUDS, with complementary data to help define density metrics, will provide better insights into the effect of environment.

\section{Observations}\label{Sct_03}

    \subsection{FAST}\label{Sct_03_01}
    
        The Five-hundred-meter Aperture Spherical Telescope (FAST, \citealp{2011IJMPD..20..989N}) has a spherical main reflector with a diameter of 500 meters, which is part of a sphere of radius 300 meters. FAST is located at longitude 106\degr51\arcmin24\farcs00074 E, latitude 25\degr39\arcmin10\farcs62653 N, Observations with zenith angles smaller than 26.4\degr can be carried out with full gain \citep{2020RAA....20...64J}. In this mode, the accessible sky area is $-0.7$\degr < Dec (J2000) $<$ 52.1\degr. FAST can also work to a maximum of zenith angle of 40\degr with lower gain and higher system temperature. The observable sky area is then enlarged to $-14.3$\degr < Dec (J2000) < 65.7\degr.
        
        During an observation,  actuators beneath the main reflector dynamically change the shape of part of the surface to form a parabola with an illuminated diameter of 300 meters. The prime focus is used to collect electromagnetic radiation across the focal plane. The design focal ratio was 0.4665, but an optimised ratio of 0.4621 is presently adopted, which gives a focal length of 138.63 meters. The position of the dome containing the primary focus and feed system is controlled by six cables, with little blockage of the optical path, resulting in low RFI scattering and low far-sidelobes compared with conventional dishes with massive support struts.
        
\begin{figure}
            \begin{center}
                \includegraphics[width=\columnwidth]{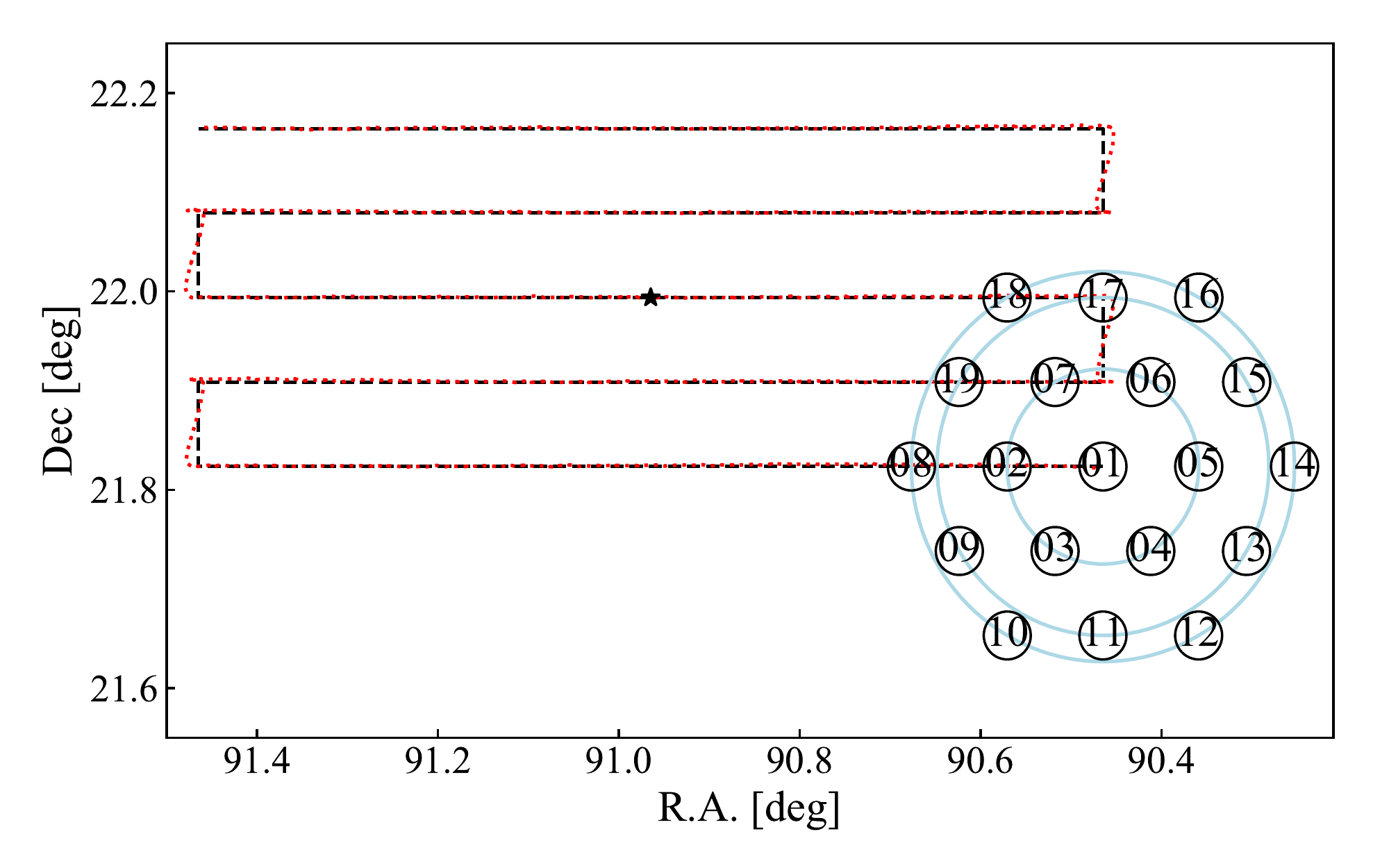}
                \caption{The geometry of the 19 beams of the FAST multibeam receiver is shown on the lower right corner. The black solid circles show the beam size of 2.9 arcmin at 1.4 GHz. The angular distance between nearby beams is 5.9 arcmin, and the spacing between nearby rows is 5.11 arcmin. The index of each beam is given in the circles. The light blue circles indicate the separation of different beams from the central beam. The black star is the position of the FUDS0 (continuum) calibrator source. The desired trace of this example calibration scan is shown with the black dashed line. The real trace of the telescope is shown with the red dotted line.
                }\label{Fig_01}
            \end{center}
        \end{figure}

        The front-end used for our observations is the 19-beam receiver \citep{2020RAA....20...64J}. According to the angular distance from centre beam, the receiver has one central beam (01), 6 beams (02 -- 07) evenly spread around an inner circle, 6 beams (09, 11, 13, 15, 17, 19) around a second concentric circle, and 6 beams (08, 10, 12, 14, 16, 18) around the third circle. The inner, second, and third circles are shown by light blue circles in Figure \ref{Fig_01}. The angular distance between the closest beams is 5.9 arcmin, and the spacing between parallel rows in the geometry shown in Figure \ref{Fig_01} is 5.11 arcmin. Each beam has two orthogonal linear polarisation channels (X and Y). The receiver can rotate to track the parallactic angle. The allowable rotation of the line joining beams 08, 02, 01, 05, 14 with a line of constant declination (current epoch) is $\pm 80$\degr. 
        
        A calibration noise diode is installed at 45\degr\ relative to two linear polarization probes in each horn. The noise diode operates at two temperatures (high at 10 K or low at 1 K) for different requirements. Since the structure of FAST main reflector depends on pointing direction, calibration of main reflector is frequently required.
        
        The FAST backend consists of 12 Re-configurable Open Architecture Computing Hardware version 2 (ROACH2) units developed by the Collaboration for Astronomy Signal Processing and Electronics Research (CASPER;  \citealp{2016JAI.....541001H, 2019SCPMA..6259502J}). Each ROACH2 unit records two pairs of polarized signals simultaneously. Hence, 10 are used to record the signals from two polarisations of 19 beams, and 2 are spare in case of breakdown. 
        
        The bandwidth of the ROACH2 is configured to be 500 MHz (frequency range 1.0 -- 1.5 GHz), split into a maximum 1,048,576 channels with a maximum frequency resolution of 476.8 Hz. For our observations, we use 65,536 channels, which gives a frequency spacing of 7.63 kHz, equivalent to a velocity spacing of 1.61 km~s$^{-1}$ in the rest frame. No Doppler shift is applied to spectra during observations. This allows us to search for \HI\ in galaxies with $cz$ in the range of $-15,908$ to $+126,034$ km~s$^{-1}$. Spectra can be recorded at maximum rate of 10 Hz, but we chose the lower rate of 1 Hz. 
        
        For each beam, three linear polarization signals (XX$^*$, YY$^*$ and XY$^*$) are recorded in {\it fits} format files, each with a maximum size of 2 GB. No Doppler shift is applied to the raw spectra before the data is recorded. At this stage, the metadata includes the position of primary focus relative to the spherical center, the skew of the lower platform, and the rotation angle of 19-beam receiver, which are all recorded in an independent spreadsheet. The sky coordinates and Doppler shifts are computed during the data reduction procedure by using the python package, {\sc astropy} \citep{2013A&A...558A..33A, 2018AJ....156..123A}. 

    \subsection{Field Selection}\label{Sct_03_02}
    
        \subsubsection{Target fields}\label{Sct_03_02_01}
            
            In order to achieve our scientific goals, we apply the following criteria to select target fields: 
            
            \begin{enumerate}
                
                \item The Declination of the target fields should be close to +25\degr\ in order for FAST to utilise its full gain for the maximum length of observing time per day.
            
                \item The target fields should be evenly distributed in RA to minimize the influence of cosmic variance and maximise the ease of scheduling night observations at any time of year.
                
                \item Target fields should avoid strong radio continuum sources ($S_\mathrm{1.4~GHz}>50$ mJy) to avoid reduction in sensitivity due to imperfect spectral subtraction. These are normally background sources, so there is no implicit selection bias. 
                
                \item The target fields should overlap with SDSS and other imaging and spectroscopic surveys to provide corresponding optical identifications and galaxy parameters. 
                
            \end{enumerate}
            
\begin{figure}
                \begin{center}
                    \includegraphics[width=\columnwidth]{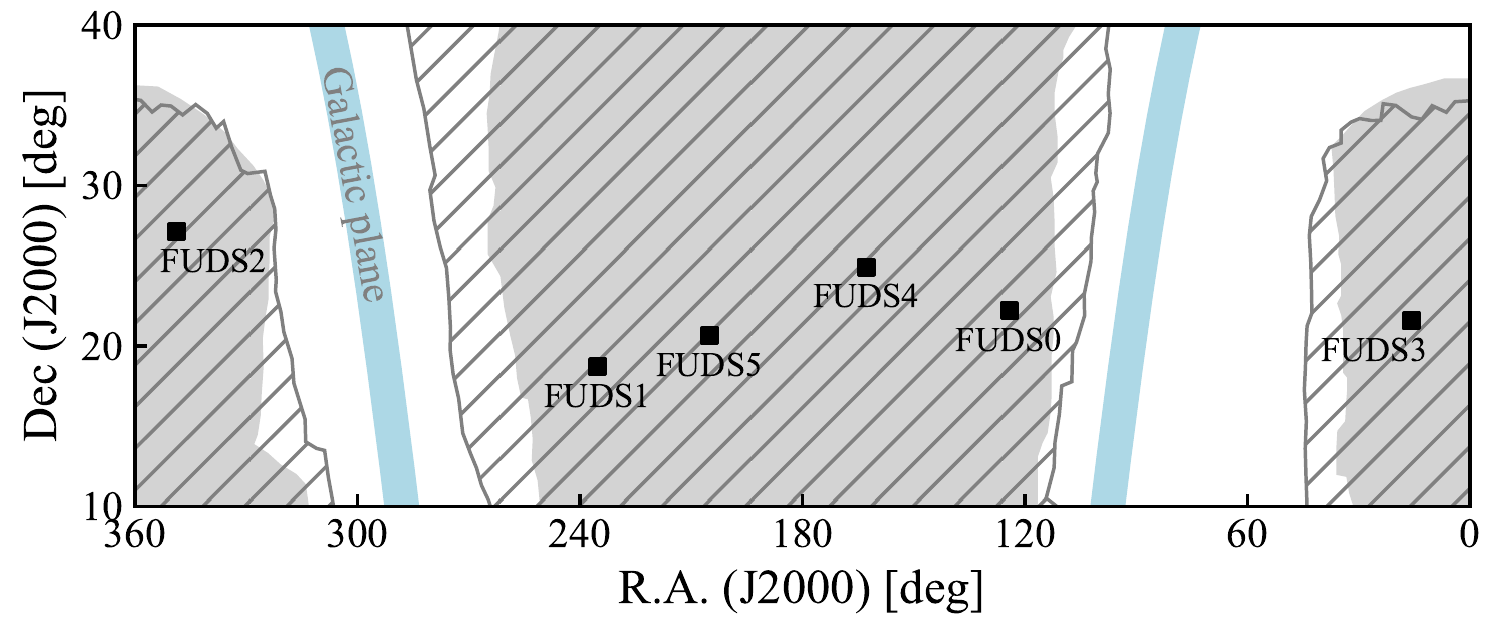}
                    \caption{This shows the six target fields in FUDS, shown by the black squares. The sky coverage of SDSS is shown with the grey area, while the DESI footprint \citep{2019BAAS...51g..57L} is shown with the hatched area. The Galactic plane is shown in light blue.}\label{Fig_02}
                \end{center}
            \end{figure}

            For the pilot FUDS0 field, we deliberately select a field which overlaps with the GAL2577 field of AUDS, which slightly compromises the continuum source criterion. In FUDS0, there are four continuum sources with flux densities larger than 50 mJy. For the remaining five target fields, all  selection criteria are met. Currently, observations for two fields are in hand: the pilot FUDS0 field, and  FUDS1. Details are listed in Table~\ref{Tab_01} and the field positions are illustrated in Figure~\ref{Fig_02}.
            
        \subsubsection{Calibrators}\label{Sct_03_02_02}
        
            Due to the deformable primary, the FAST gain barely changes at different zenith angles (within 26\fdg4) \citep{2020RAA....20...64J}. We only use one flux calibrator for each target field in FUDS using the following selection criteria:
    
            \begin{enumerate}
        
                \item Strong ($> 1.0$ Jy at 1.4 GHz) and stable flux density (marked as "P" in the VLA calibrator list\footnote{https://science.nrao.edu/facilities/vla/observing/callist} for the VLA D configuration, which has the beam size most similar to FAST).
        
                \item In order to avoid confusion, no other strong continuum sources close to the selected calibrator are listed in the NVSS catalogue \citep{1998AJ....115.1693C}.
        
                \item Calibrators are close to target fields, and can be observed within a zenith angle of +26\fdg4.
        
            \end{enumerate}
    
            Flux densities at frequencies near 1.0 -- 1.5 GHz and their uncertainties were extracted from NED\footnote{http://ned.ipac.caltech.edu/}. To derive the frequency dependence of the calibrator flux densities, we use a power law to fit the the measurements and uncertainties by using an optimised $\chi^2$ method:
            \begin{equation}
                S_{\nu} = S_{\nu_0} \left(\frac{\nu}{\nu_0}\right)^{\alpha},
                \label{Equ_01}
            \end{equation}
            where the $\nu$ is frequency, $\nu_0$ is the reference frequency (1.4 GHz), and $\alpha$ is the spectral index. For FUDS, we assume the variability of compact sources is negligible during the course of our observations of a particular field.
    
\begin{figure}
                \begin{center}
                    \includegraphics[width=\columnwidth]{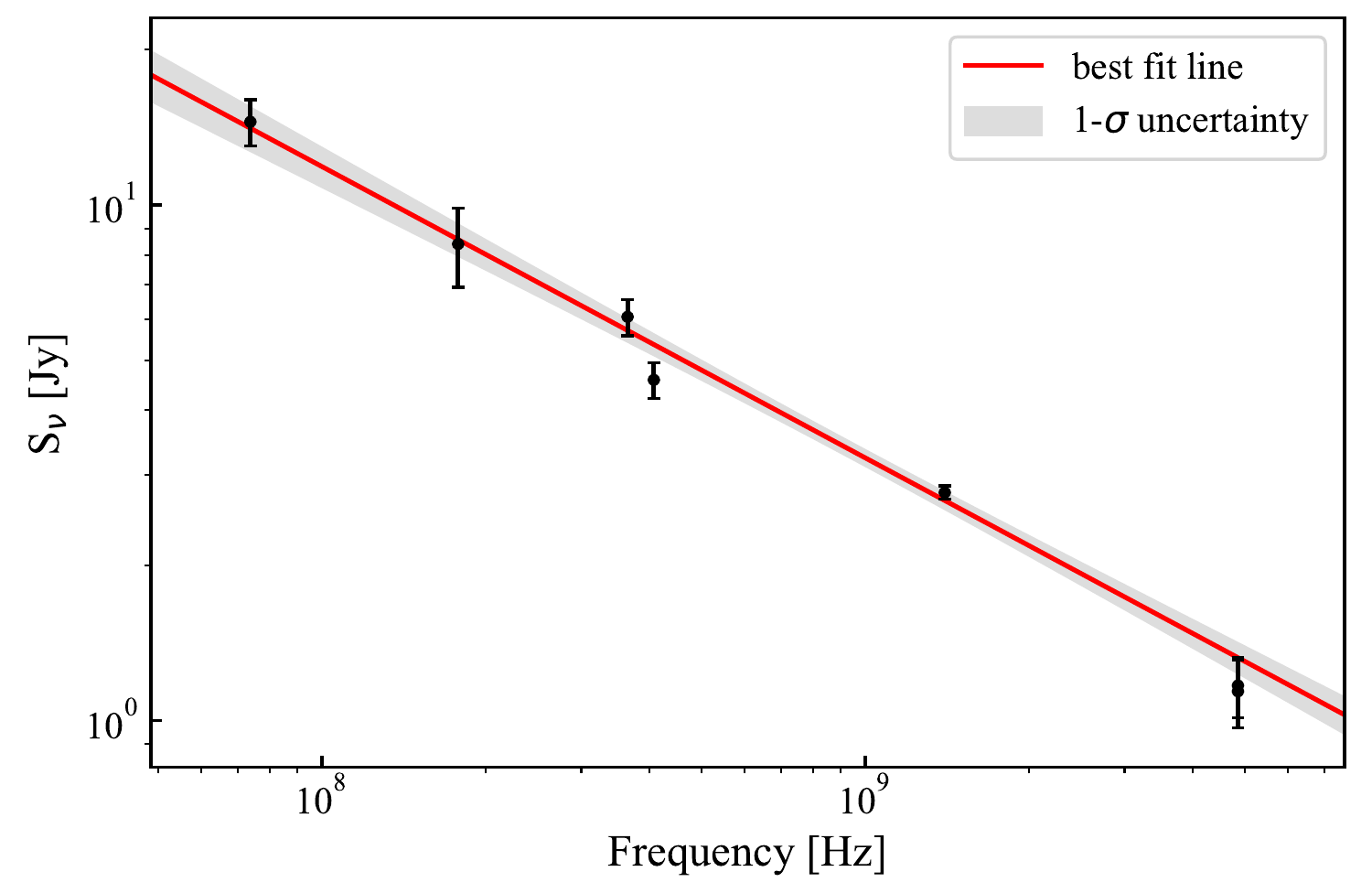}
                    \caption{Flux densities and corresponding uncertainties for the calibrator at different frequency from previous works. The red line is the best fit of Equation \ref{Equ_01}. The 1-$\sigma$ uncertainty is shown by grey shadow.}\label{Fig_03}
                \end{center}
            \end{figure}

            As an example, we show the parameters for the FUDS0 calibrator, 0603+219, in Table~\ref{Tab_02} and show the fit to its continuum spectrum in Figure~\ref{Fig_03}. The VLA flux density is 2.75 Jy at 1.4 GHz and the computed spectral index is $\alpha=0.565\pm 0.034$. For computing frequency-dependent system parameters, we anchor the flux density with the accurate VLA measurement, and use our computed spectral index.
    
            \begin{table*}
                \centering
                \caption{The properties of FUDS0 calibrator. The first four rows show the properties of the calibrator from VLA calibrator list. The flux density at different frequency is listed from the fifth to eleventh row. The parameters of Equation \ref{Equ_01} from best fit are given in Row 12 and 13.}
                \label{Tab_02}
                \begin{tabular}{lrr}
                    \hline
                    \hline
                    Property      & Value              & Ref \\
                    \hline
                    Name          & 0603+219           & VLA calibrator list\\
                    R.A. (J2000)  & 06$^{\rm h}$03$^{\rm m}$51\fs557091   & VLA calibrator list\\
                    Dec (J2000)   & +21\degr59\arcmin37\farcs697500   & VLA calibrator list\\
                    $S$(1.4 GHz)  & 2.75 (-) Jy        & VLA calibrator list\\
                    $S$(73.8 MHz) & 14.5 (1.49) Jy     & \citet{2007AJ....134.1245C} \\
                    $S$(178 MHz)  & 8.4 (1.47) Jy      & \citet{1965MmRAS..69..183P} \\
                    $S$(365 MHz)  & 6.07 (0.49) Jy     & \citet{1996AJ....111.1945D} \\
                    $S$(408 MHz)  & 4.58 (0.376) Jy    & \citet{1974AnAS...18..147F} \\
                    $S$(1.4 GHz)  & 2.77 (0.0832) Jy   & \citet{1998AJ....115.1693C} \\
                    $S$(4.85 GHz) & 1.168 (0.157) Jy   & \citet{1991ApJS...75.1011G} \\
                    $S$(4.85 GHz) & 1.143 (0.171) Jy   & \citet{1991ApJS...75....1B} \\
                    $S_{\nu_0}$   & 2.67(0.09) & best fit \\
                    $\alpha$      & -0.565(0.034) & best fit \\
                    \hline
                    \hline
                \end{tabular}
            \end{table*}
    
    \subsection{Observational strategy}\label{Sct_03_03}

        \subsubsection{Target fields}\label{Sct_03_03_01}
            
\begin{figure}
            \begin{center}
                \includegraphics[width=\columnwidth]{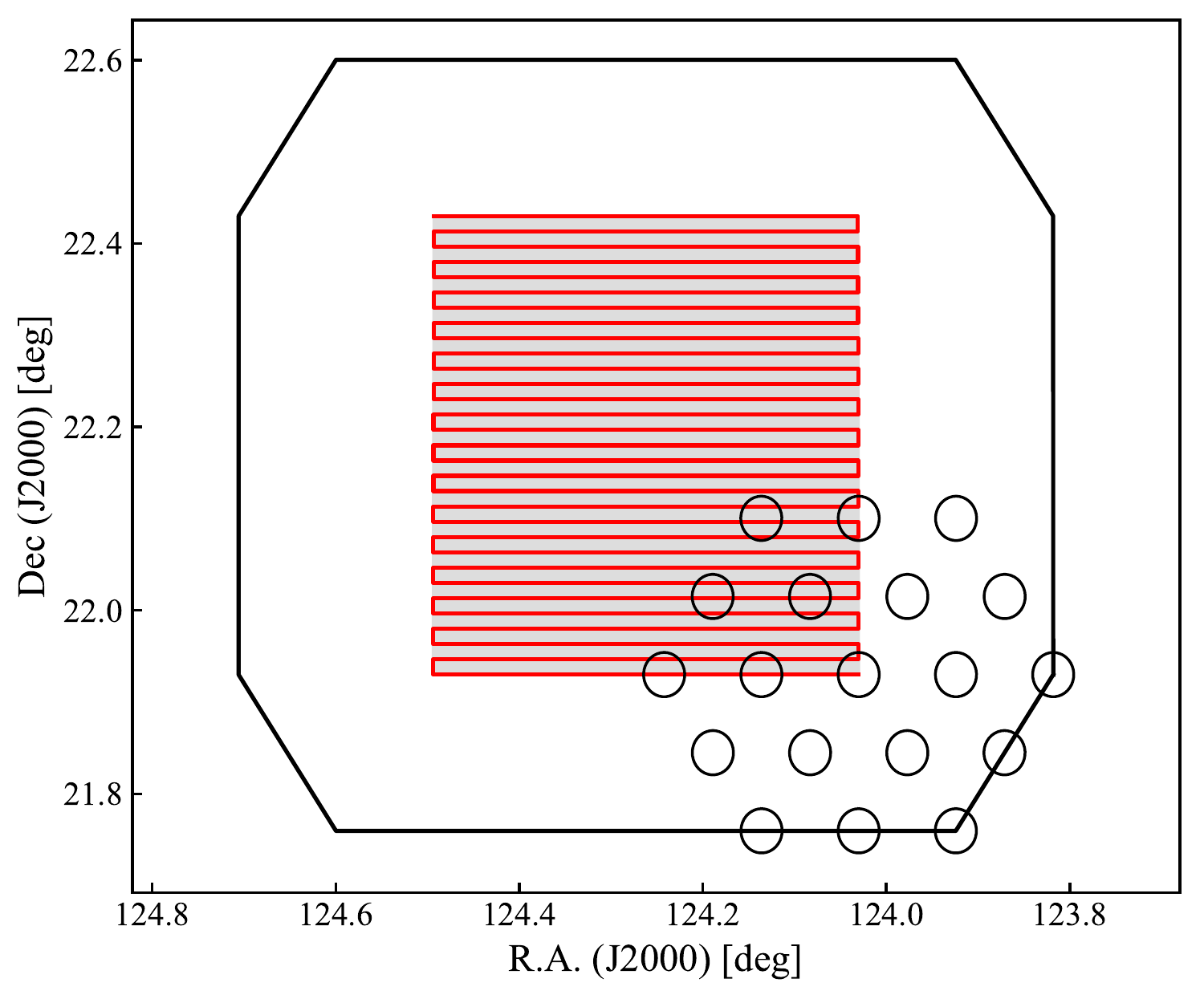}
                \caption{The position of the central beam whilst scanning the FUDS0 target field is shown by the red lines. The grey area has high sensitivity, but coverage extends to the outer black lines. The start positions for each of the 19 beams are given by the black circles.}
                \label{Fig_04}
            \end{center}
        \end{figure}

            As shown in Figure~\ref{Fig_04} for the FUDS0 field, an on-the-fly (OTF) observation mode is employed. Each target field is $30'\times30'$, as scanned by the central beam, with a rotation angle of 0\degr. However, due to the multiple beam pattern, this results in a field size of $50'\times50'$ with good sensitivity. The spacing between scans is 1 arcmin, which is smaller than the Nyquist sampling interval for an individual 2.9 arcmin beam at 1.4 GHz. The target field is scanned in both RA and Dec direction with a rate of 15 arcsec~s$^{-1}$. Each scan takes 120 s and spectra are recorded every second. We use the high temperature calibration noise diode (10 K) to inject noise for 1 s every 60 s. The total observing time per set of 31 scans is around 1.2 hrs (62 min scanning, 10 min overhead), or 2.4 hrs for the combined RA and Dec scans. Multiple sets of similar observations are then combined into the final data cube.
            
    \subsubsection{Calibrators}\label{Sct_03_03_02}
     
            The normal FAST calibration procedure is to point each beam in turn at a calibration source for $\sim 1$ min. However \citet{2020RAA....20...64J} use an alternative technique of scanning across calibrators which has the advantage of also measuring beam size and pointing offsets. We take this technique, but extend it to the full range of frequencies from 1.0 to 1.5 GHz, as FUDS has good sensitivity to high-redshift galaxies. As mentioned before, OTF mode is used to scan the calibrator with a receiver rotation angle of 0\degr, but along scan lines of 1\degr\ in length (see Figure~\ref{Fig_01}), which allows all the beams in the same row pass the calibrator in a single scan with adequate baseline either side. The spacing between scan lines is 5.11 arcmin.
            
            Since the rotation angle is relative to the Dec at the current epoch, a slight tilt in the scan is introduced except for the central beam. The maximum deviation is $\sim$1.5 arcsec for the outer beams and $\sim$0.7 arcsec for the inner beams, which is much smaller than  the pointing accuracy ($\sim 8$ arcsec). As for the target observations, the scan speed is 15 arcsec~s$^{-1}$, and the spectra are recorded every 1 sec. The high temperature 10 K noise diode is injected in a square wave pattern with an alternating pattern of 1 sec on and 1 sec off. The total observation time is around 30 min (20 min scanning, 10 min overhead).
        
\section{Radio frequency interference (RFI)}\label{Sct_04}
    
    RFI reduces the signal-to-noise ratio of observational data. A list of RFI-affected frequencies is useful for pre-flagging when reducing target field data. As a newly built radio telescope, the FAST list is yet incomplete. From the initial FUDS data, we identify four types of RFI: 
    
    \begin{itemize}
    
        \item Signals from the global navigation satellite system (GNSS), which include Galileo, GLONASS, GPS, and Beidou. The intensity of these signals depends on the number of satellites in the sky.  Within our observational window, they transmit in the frequency range 1.15 to 1.30 GHz. 
            
        \item Signals from synchronous communication satellite, AsiaStar. These signals are strong and range from 1.48 to 1.49 GHz.
        
        \item Signals from the civil aviation radar. They occur at 1.03 and 1.09 GHz. An occasional strong pulse radar signal appears at 1.105 and 1.131 GHz.
        
        \item Signals generated by the refrigerating dewar in the compressor. These cover the entire 500 MHz bandwidth. The signals have a temperature of a few K, and a linewidth of about 1 MHz. In order to solve the issue, the compressor was screened in Aug 2021. Raw FUDS data taken since then do not show this RFI, even in spectra averaged over 20 min. However, initial FUDS0 and some FUDS1 data are contaminated.
    
    \end{itemize}
        
    \subsection{RFI detection and baseline fitting}\label{Sct_04_01}
            
        The GNSS signals are strong (having comparable intensity of the system temperature of FAST) and highly variable in time. They also occupy a large fraction of the FAST bandwidth (37.5\%). The signals can affect the system temperature and its measurement even at non-RFI frequencies. The RFI from the compressor also adds uncertainty for flux intensity measurements by biasing the bandpass calibration. In order to mitigate these affect, we identify the RFI and to fit an underlying spectral baseline using the following steps:
        
        \begin{enumerate}
        
            \item The spectrum is divided into several segments with an overlap of 15 MHz. The width of each segment varies from 90 MHz to 240 MHz, depending on the number of channels that are free of RFI.
            
            \item For each segment, a Gaussian weighted polynomial function is employed to fit the baseline. The standard deviation of the Gaussian weight is 0.17 times the segment width. The order of the fit is iterated from 0 to 6. For each order, a pseudo-RMS is calculated using 1.4826$\times$MAD (Median Absolute Deviation). The channels with absolute values exceeding $7\times$ RMS and immediate neighbouring channels with absolute values exceeding $3\times$ RMS are marked as RFI.
            
            \item A temporary baseline for the whole spectrum is determined by combining the baselines of each segment with Gaussian weights.
                
            \item The above procedures are iterated on a smoothed baseline-removed spectrum with increasing smoothing length (25, 50, 100 and 200 kHz). The intermediate baseline is generated by summing together the temporary baselines from each iteration. 
                
            \item The final baseline is the intermediate baseline generated in the final iteration. All the channels ever marked as RFI in each iteration are recorded as RFI in the final spectrum.
            
        \end{enumerate}
        
        \begin{figure}
            \begin{center}
                \includegraphics[width=\columnwidth]{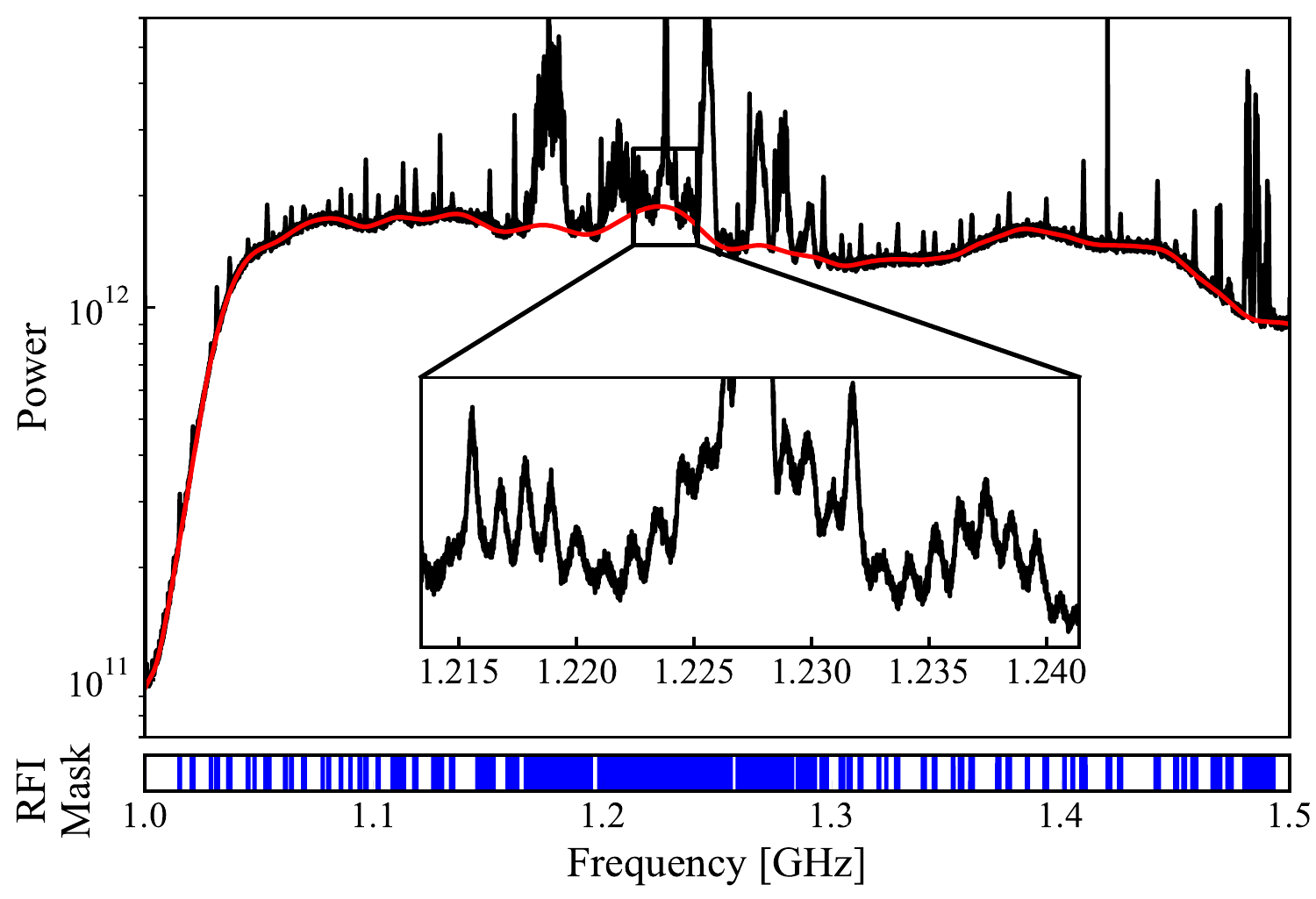}
                \caption{The upper panel shows a total power spectrum (polarisation  XX$^*$) from the central beam (black line) and the underlying fitted baseline (red line). The units are arbitrary. The spectrum around 1.23 GHz is zoomed in to show the standing wave. The frequencies of the detected RFI are indicated by the blue bars in the lower panel.}
                \label{Fig_05}
            \end{center}
        \end{figure}
            
        We note that RFI flagging in raw data is time consuming. Since the RFI is stable in frequency in a short period (around 20 min, see Section \ref{Sct_04_02}), we only perform this procedure on one spectrum per 60 spectra (the same frequency as the noise injection in the target fields). The template for flagged RFI is therefore used for the next 59 spectra.
        
        In Figure \ref{Fig_05}, we show a total power spectrum  (XX$^*$ polarisation) from a single 1~s integration from the FUDS0 central beam. The fitted baseline and the identified regions with RFI are demonstrated. About 41.7\% of the spectrum contains RFI at some level. This procedure may miss weak RFI, but the baseline fitting is relatively robust.
        
    \subsection{Compressor RFI}\label{Sct_04_02}
            
\begin{figure}
            \begin{center}
                \includegraphics[width=\columnwidth]{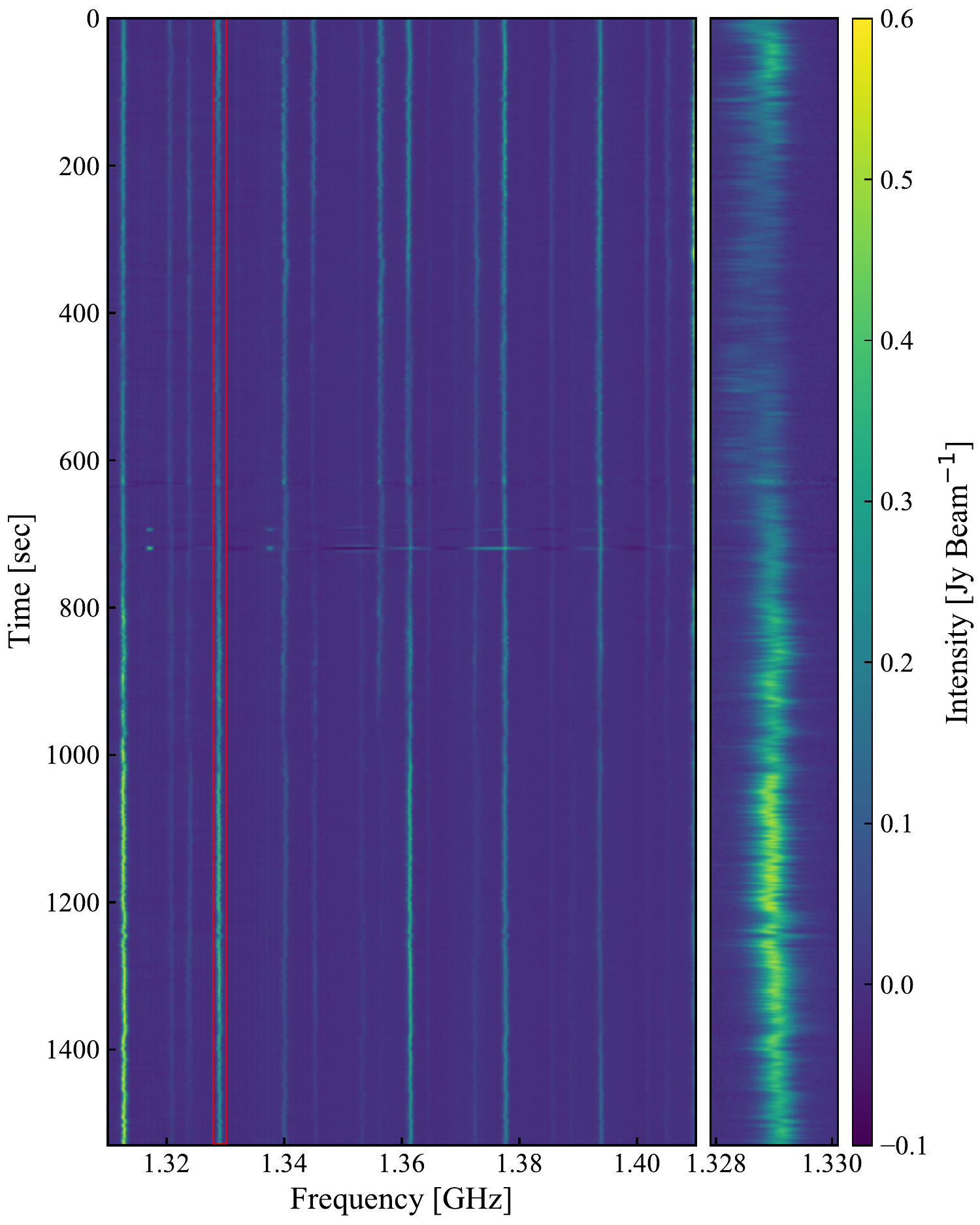}
                \caption{In the left panel, calibrated data between 1.31 and 1.41 GHz is shown for the XX$^*$ polarization channel of beam 1 as a function of frequency (in GHz) and time (in sec). The right panel is a zoom (in frequency space) of the region indicated by the red rectangle. The signals present are due to RFI from a compressor in FAST data taken prior to July 2021.}
                \label{Fig_06}
            \end{center}
        \end{figure}

\begin{figure}
            \begin{center}
                \includegraphics[width=\columnwidth]{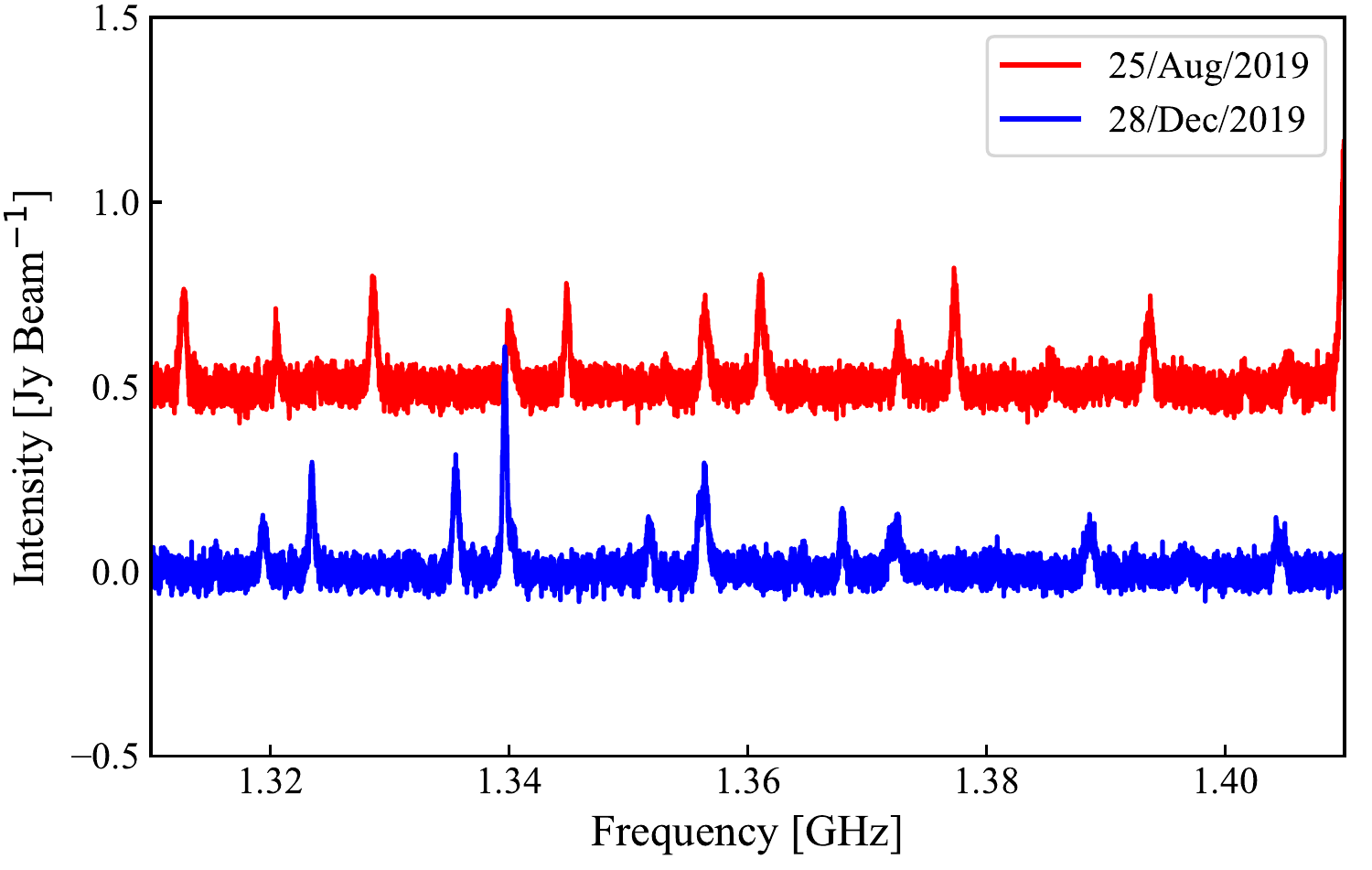}
                \caption{The spectra of one polarization (XX$^*$) from beam 1 on 2019 Aug 25 (red) and 2019 Dec 28 (blue). The former spectrum has been vertically shifted by 0.5 Jy beam$^{-1}$ for clarity.}
                \label{Fig_07}
            \end{center}
        \end{figure}

        To characterise the signals generated by the refrigerating dewar in the compressor, we analyse the calibrator data taken on 2019 August 25. In Figure~\ref{Fig_06}, we show a waterfall image of the calibrated spectra between frequencies of 1.31 and 1.41 GHz. This frequency range excludes the GNSS RFI and \HI\ emission from the Milky Way. The compressor signals (the vertical stripes) generally have an intensity of about 0.3 Jy and a linewidth of $\sim 1$ MHz. The signals are variable in amplitude and frequency (middle panel of Figure~\ref{Fig_06}) and also shift in frequency when comparing data on different days (see Figure \ref{Fig_07}). The signals do not depend on azimuth or elevation as the compressor is housed in the focus cabin.
    
\begin{figure*}
            \begin{center}
                \includegraphics[width=0.45\columnwidth]{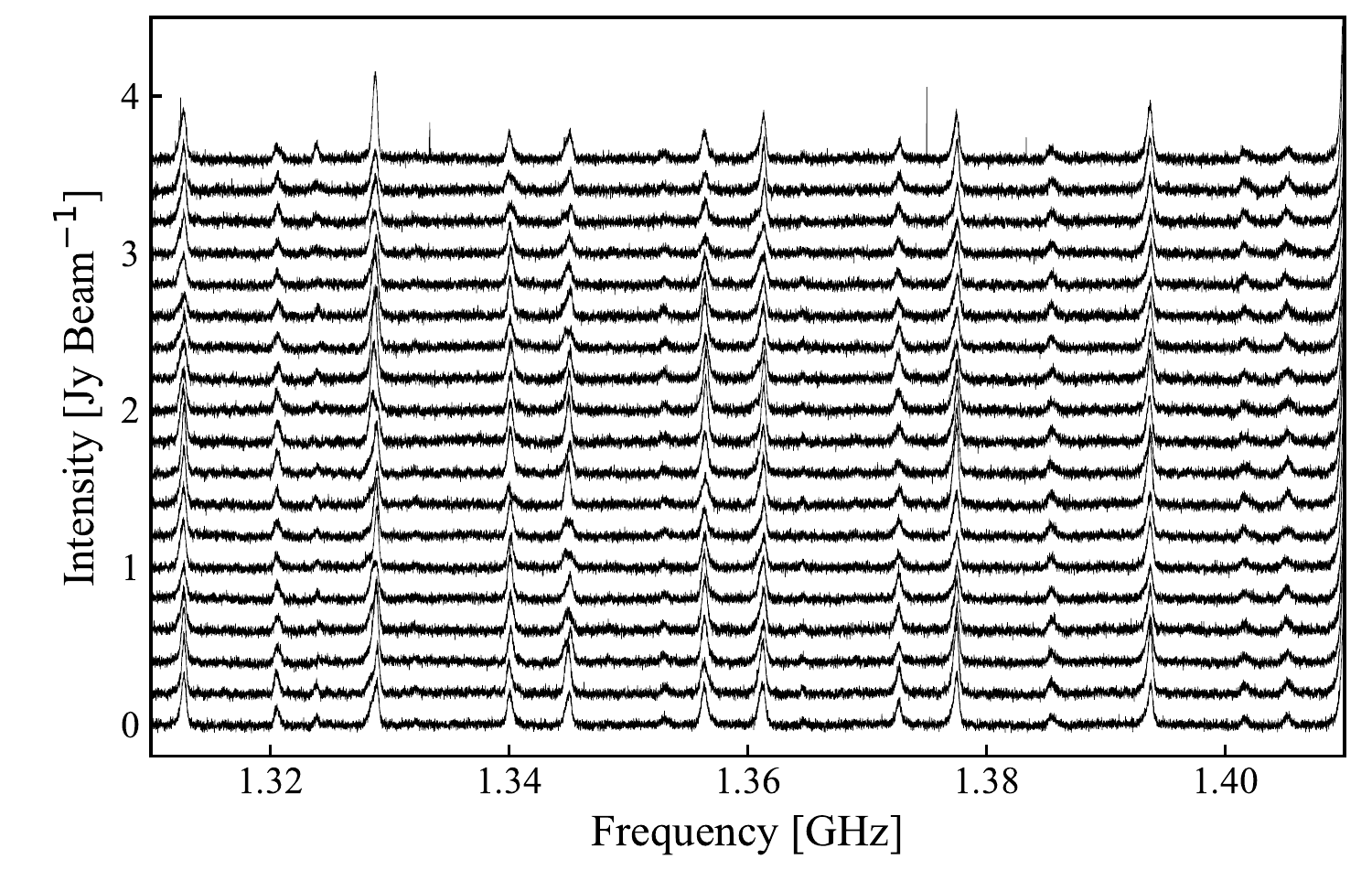}
                \includegraphics[width=0.45\columnwidth]{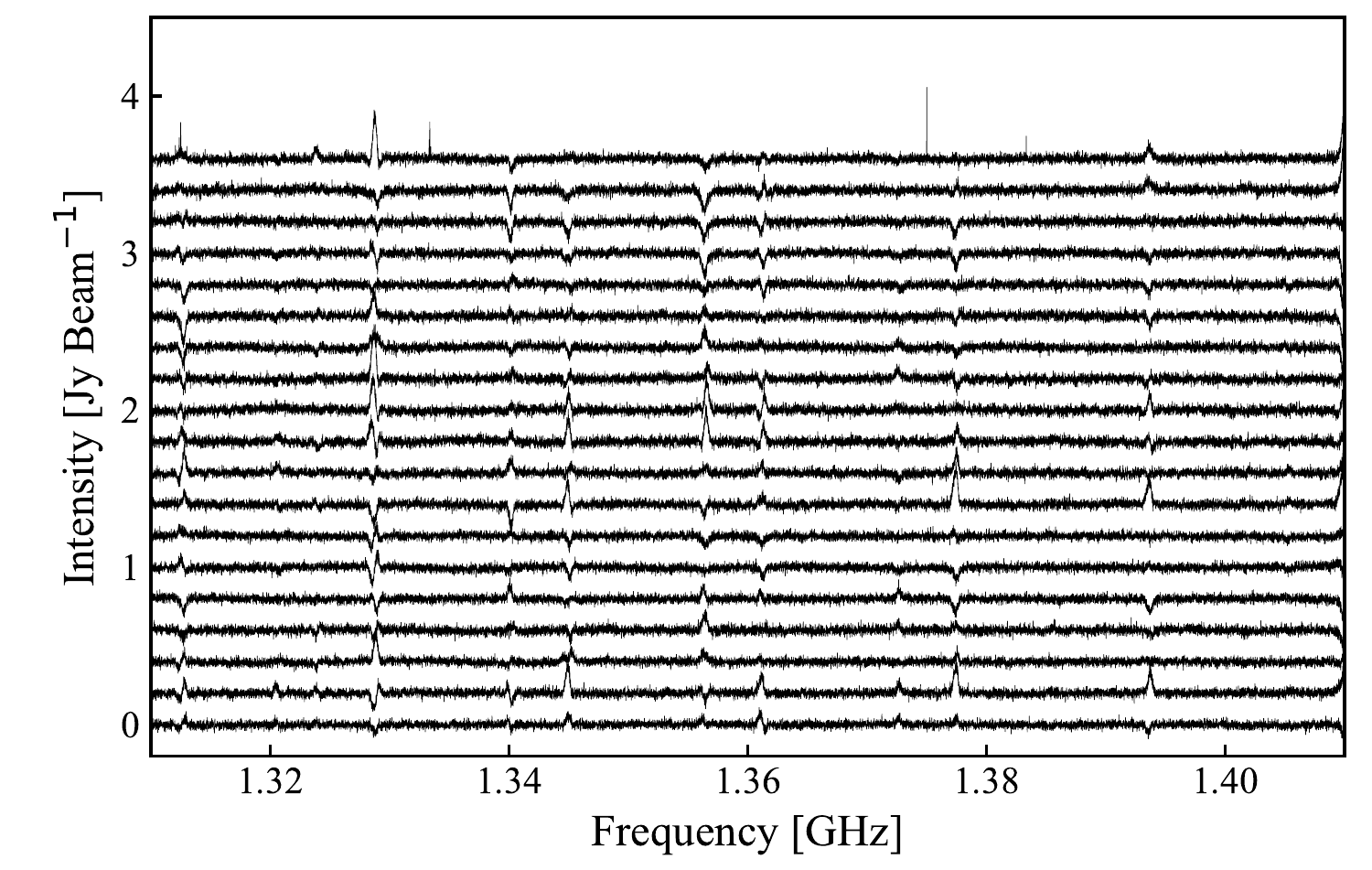}
                \caption{Example spectra of all beams observed at one time (polarisation XX$^*$ from beams 1 to 19 are offset from the bottom to the top) are shown in the left panel. The frequency and intensity of the local RFI is fairly constant across all beams. The right panel shows the residual spectra after removing a template constructed from all beams. The RFI intensity has been reduced.}
                \label{Fig_08}
            \end{center}
        \end{figure*}

        Calibrated spectra of all 19 beams from the same observation are shown in the left panel of Figure \ref{Fig_08}. There is a similarity in frequency and intensity of the compressor RFI across all 19 beams. We create a smoothed median spectrum from all beams and use it as template to subtract from all spectra. The residual spectra after RFI removal are shown in the right panel of Figure \ref{Fig_08}. Removal of the compressor signals is only performed after the calibration due to large power differences between the beams. Large sources which extend across several beams will potentially have their flux density reduced. We note that further flagging is required to remove residual RFI.
    
\section{Calibration}\label{Sct_05}

    \subsection{Gain stability}\label{Sct_05_01}
    
        The FAST calibration method involves switching on and off a noise diode, which is assumed to have a constant amplitude. Once the spectrum of noise diode has been calibrated against a radio continuum source, the diode noise can then be used to measure the system equivalent flux density (SEFD) in Jy, $S_\mathrm{sys}$ and its stability. Because the high temperature noise diode elevates $T_{\rm sys}$ by nearly 50\%, it is impractical to use with a 50\% duty cycle for target field observations. We therefore needed to confirm that the receiver calibration was stable in a mode where the calibration `on' spectrum was only taken for 1 sec every 60 sec.
        
\begin{figure}
            \begin{center}
                \includegraphics[width=\columnwidth]{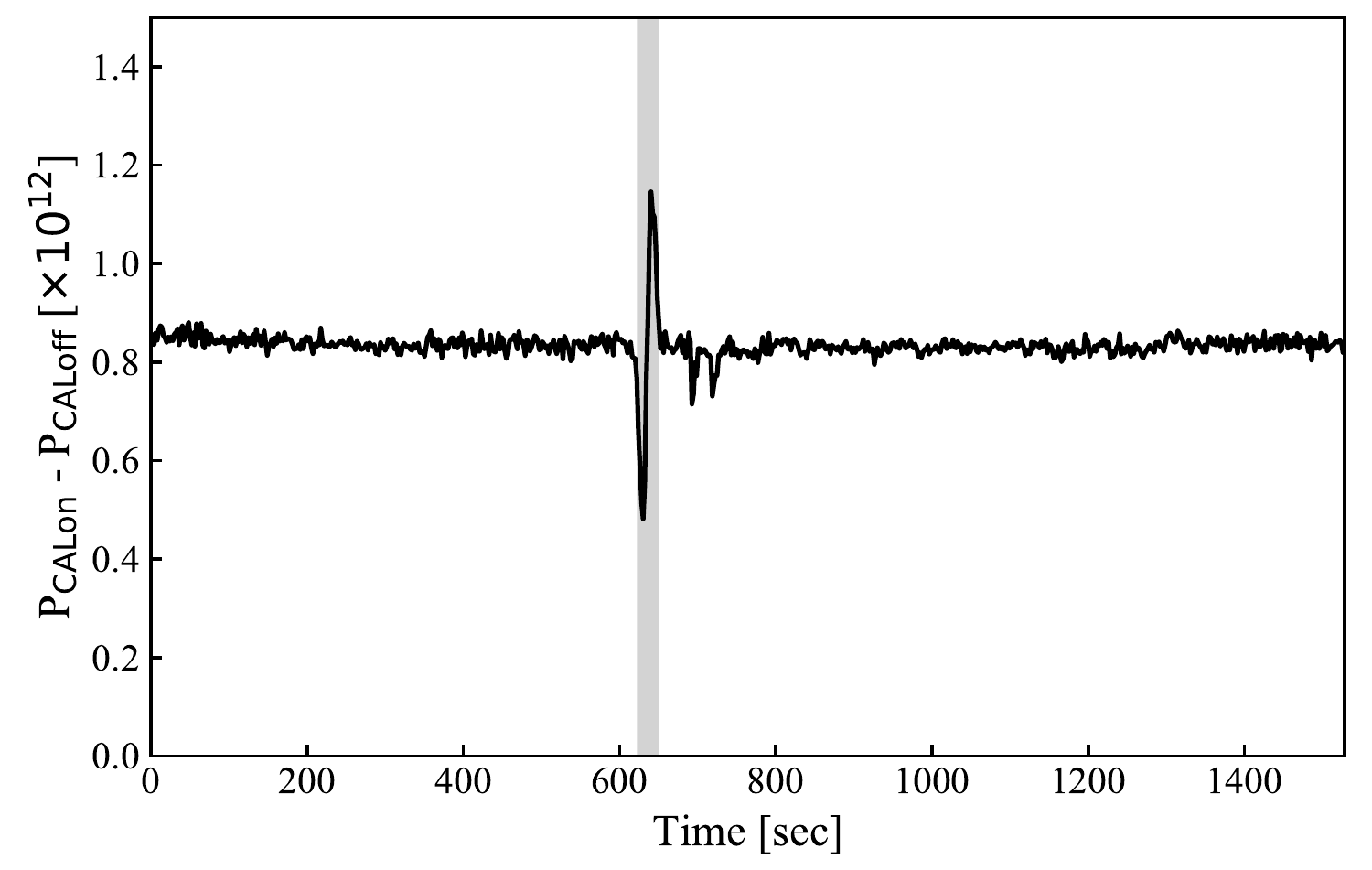}
                \caption{The power difference between successive integration cycles with the calibration noise diode on and off, respectively, as a function of time. The power is calculated after removing edge channels and channels affected by RFI. The units of power are arbitrary. The grey area indicates a time range where the calibrator distance was less than 3 arcmin.}
                \label{Fig_09}
            \end{center}
        \end{figure}

        To confirm this, we first removed the identified RFI channels (see Section \ref{Sct_04_01}), and channels that are outside of the official bandpass ($<1.05$, or $>1.45$ GHz). The noise diode was switched on and off at a rate of 1 Hz while performing a calibrator observation, and the difference in the mean power in each pair of on/off spectra is calculated. Figure \ref{Fig_09} shows the result, which indicates that no significant variation  occurs over any time frame less than 1,500 sec. This confirms that both the noise diode and the receiver gain are adequately constant within this period.
    
    \subsection{Instrumental characterisation}\label{Sct_05_02}
    
        For a radio telescope, a Gaussian function is a good approximation for angular response of the beam (the point spread function). The following function can therefore be used to fit a 1-dimensional scan of a calibrator where the noise diode (CAL) is either on or off:
        \begin{equation}
            f(d)=
            \left\{ \begin{array}{ll}
                Ae^{-\frac{(d-d_o)^2}{2\sigma^2}}+C_1 &\text{(CAL on)}\\
                Ae^{-\frac{(d-d_o)^2}{2\sigma^2}}+C_2 &\text{(CAL off)}
            \end{array} \right.
            \label{Equ_02}
        \end{equation}
        where $d$ is the angular distance offset from the calibrator, $d_o$ is the calibrator position, $A$ is the amplitude, $\sigma$ is the standard deviation, and $C_1$ and $C_2$ are the CAL on and off baselines, respectively. The flux density of the noise diode is then $S_\mathrm{noise}=\frac{C_1-C_2}{A}\times S_\mathrm{calibrator}$, the SEFD $S_\mathrm{sys}=\frac{C_2}{A}\times S_\mathrm{calibrator}$, and the full width half power (FWHP) beamwidth is $\theta=\sqrt{8\ln2} \sigma$, where $S_\mathrm{calibrator}$ is the flux density of the calibrator. We also measure the pointing error, $\sigma_\mathrm{pointing}$ from the RMS of the deviation of $d_o$ from the catalogued position.
        
        We firstly calculate mean instrumental parameters, $\bar{S}_\mathrm{noise}$, $\bar{S}_\mathrm{sys}$, $\bar{\sigma}_\mathrm{pointing}$, and $\bar{\theta}$ averaged over the frequency range 1.05 to 1.45 GHz, using the VLA calibrator flux density at 1.4 GHz. 
        
\begin{figure}
            \begin{center}
                \includegraphics[width=\columnwidth]{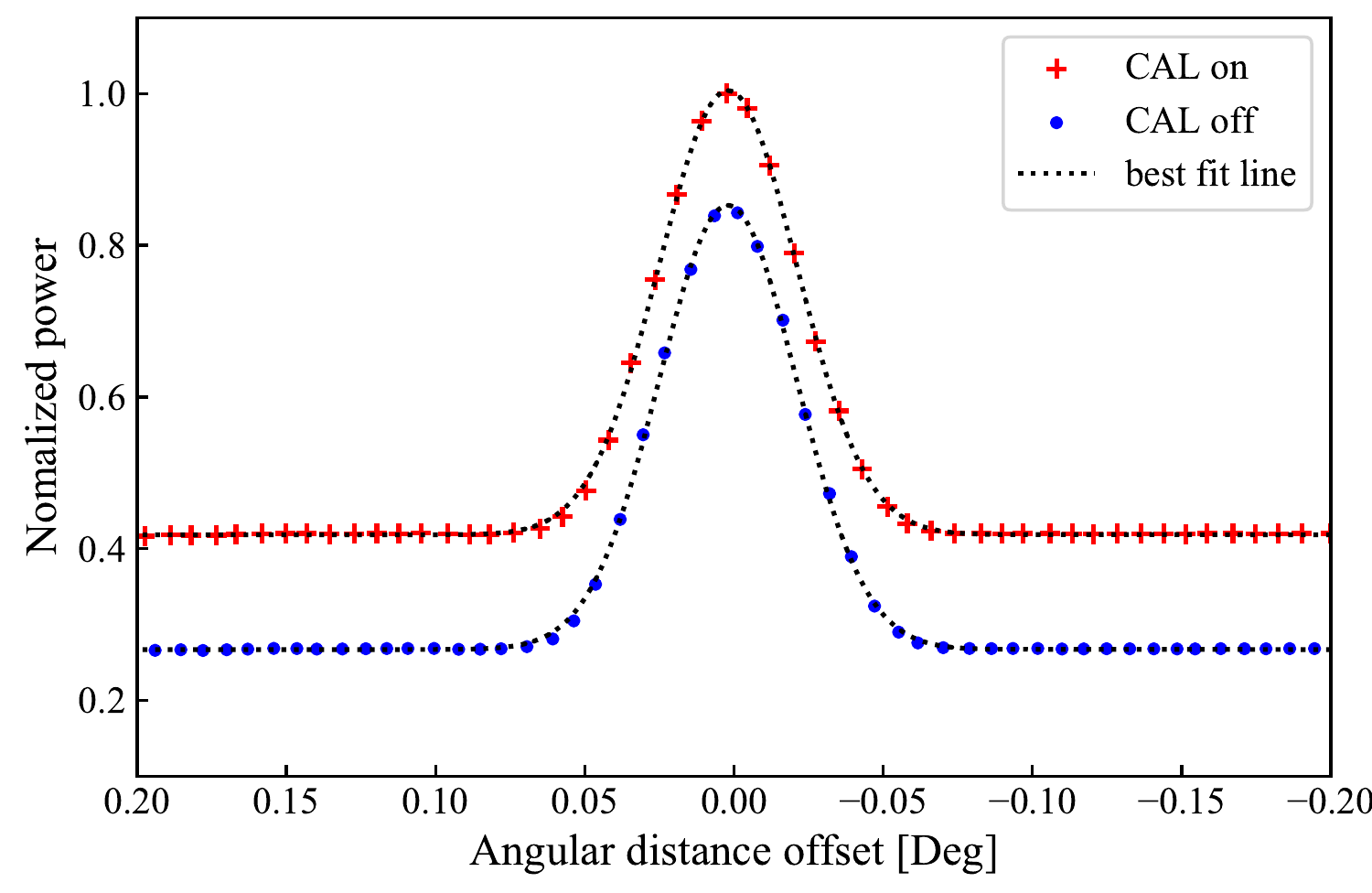}
                \caption{The mean power (in arbitrary units) of polarisation XX$^*$ from beam 1 as a function of angular distance from the calibrator. Channels containing RFI and edge channels are removed prior to calculating the mean. The mean power for integration cycles with the noise diode on and off are shown with red plus signs and blue dots, respectively. The best fits curves using Equation \ref{Equ_02} are shown.}\label{Fig_10}
            \end{center}
        \end{figure}

\begin{figure*}
            \begin{center}
                \includegraphics[width=0.45\columnwidth]{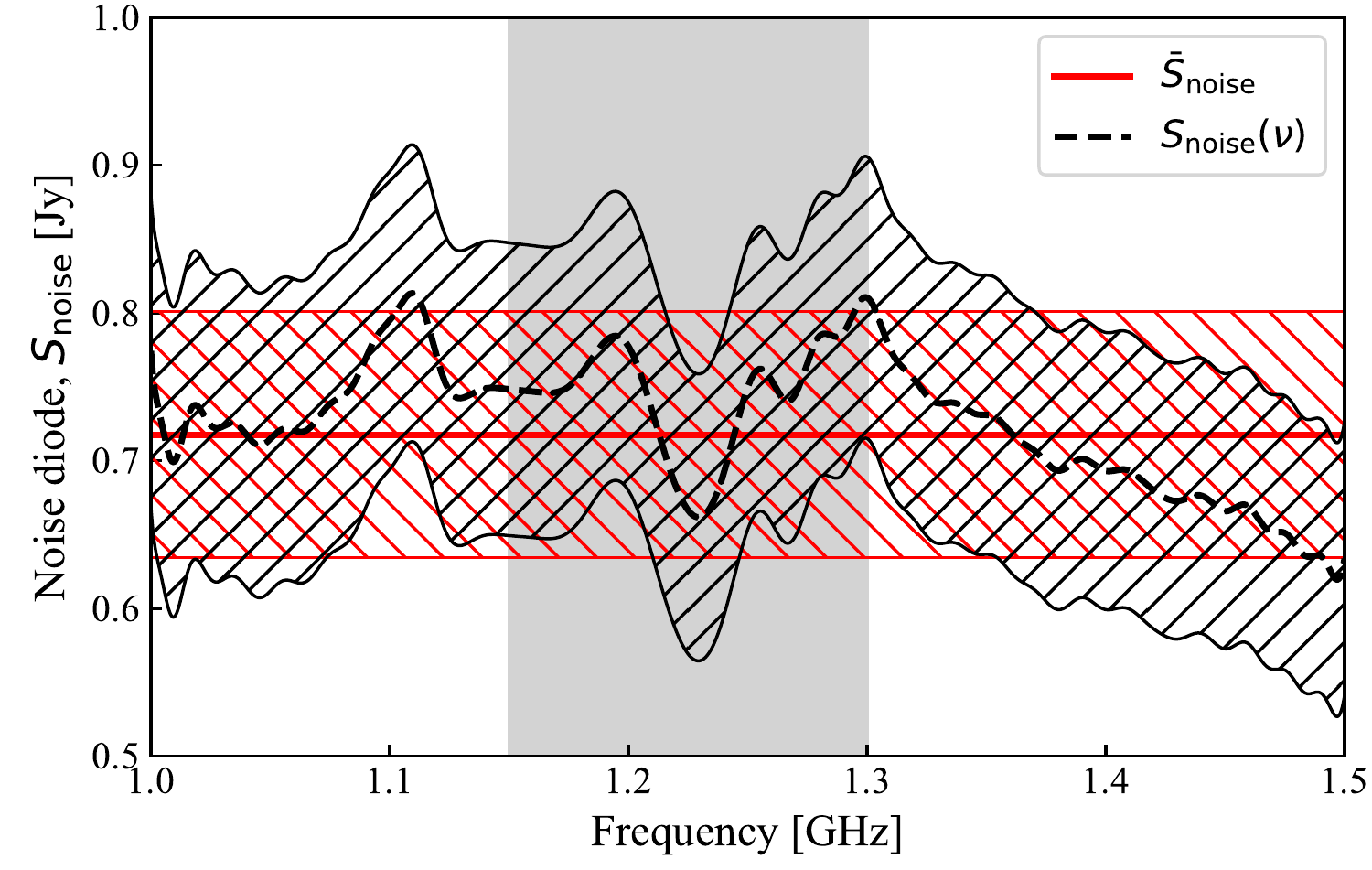}
                \includegraphics[width=0.45\columnwidth]{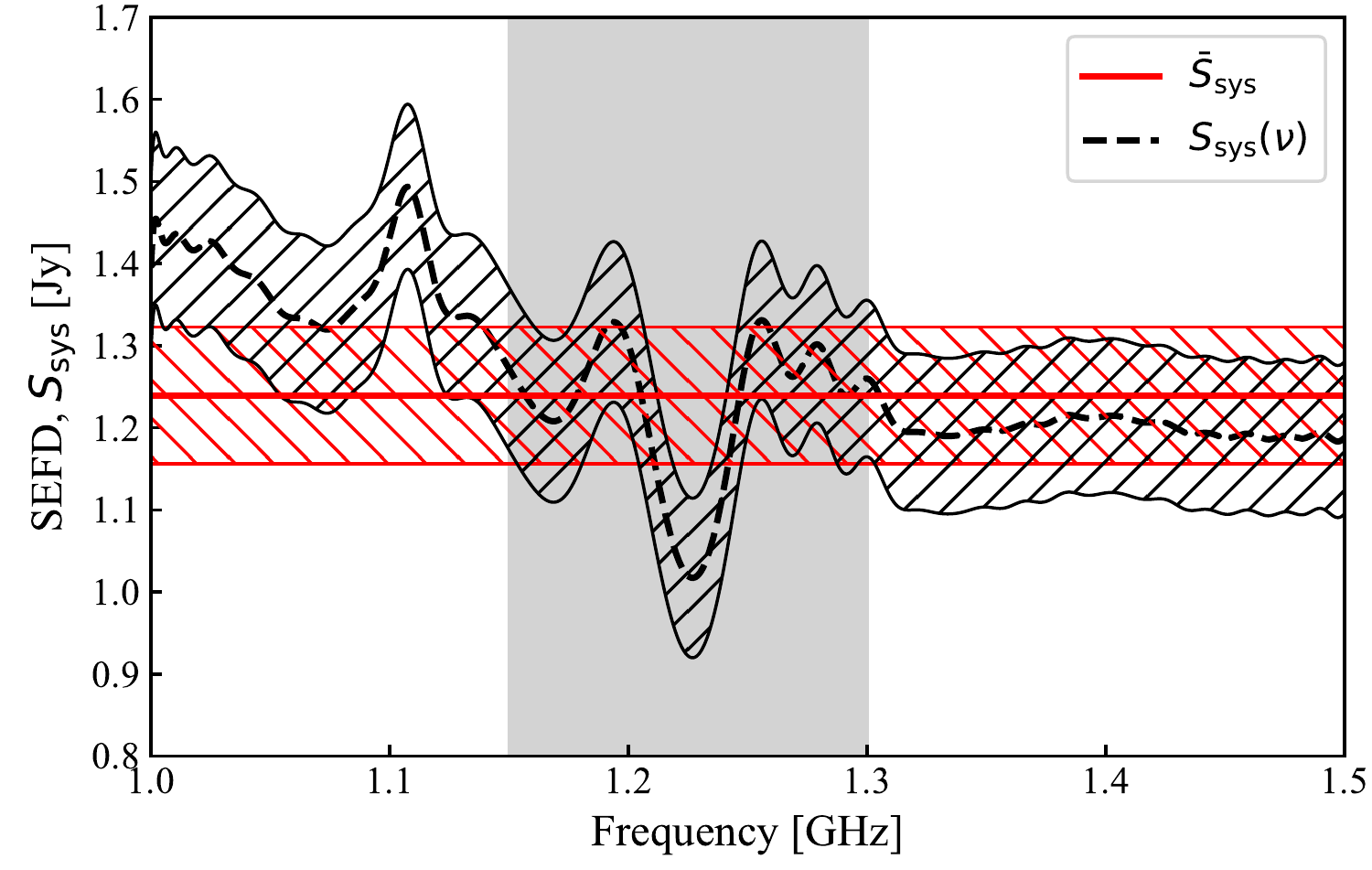}
                \includegraphics[width=0.45\columnwidth]{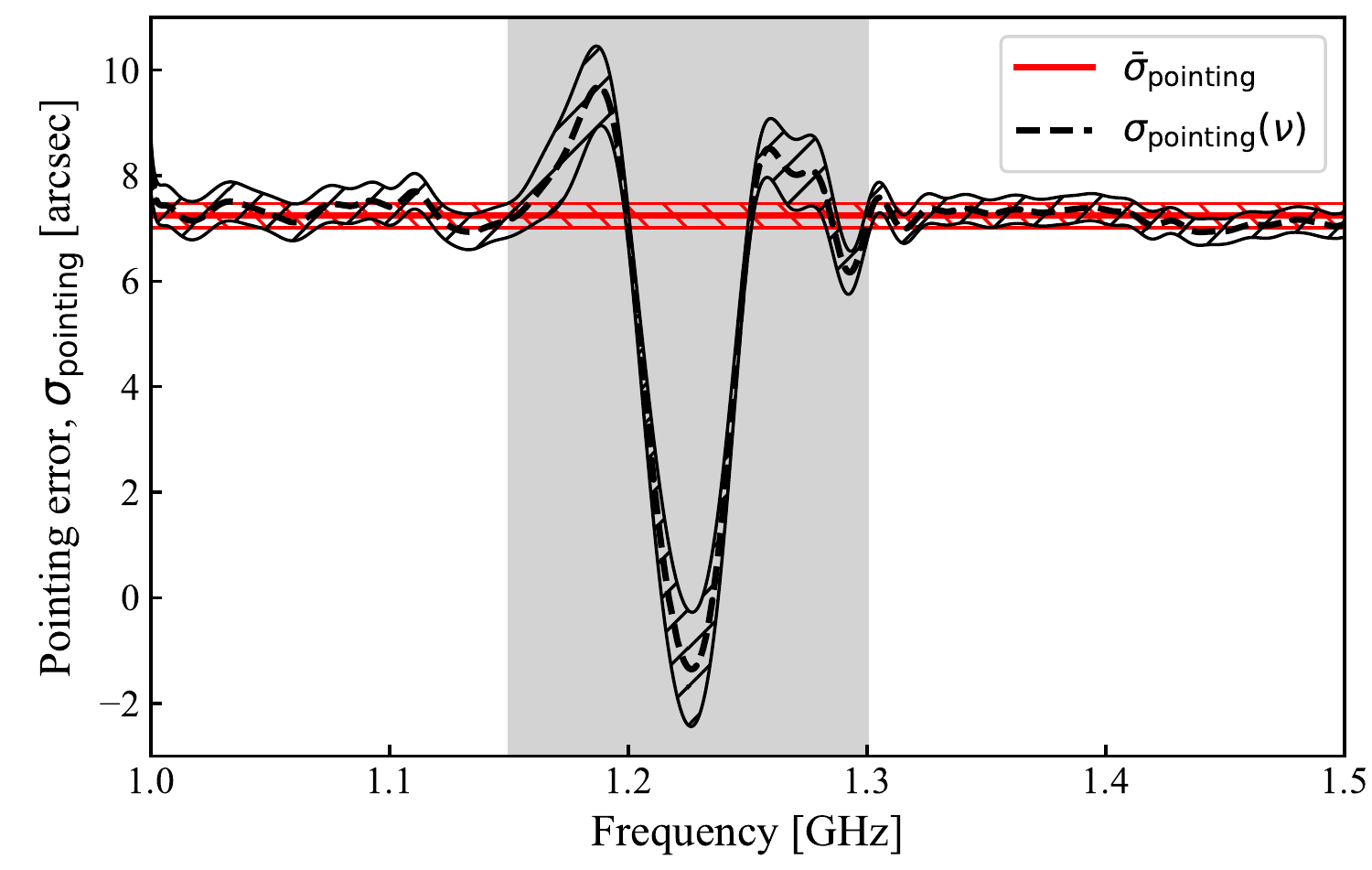}
                \includegraphics[width=0.45\columnwidth]{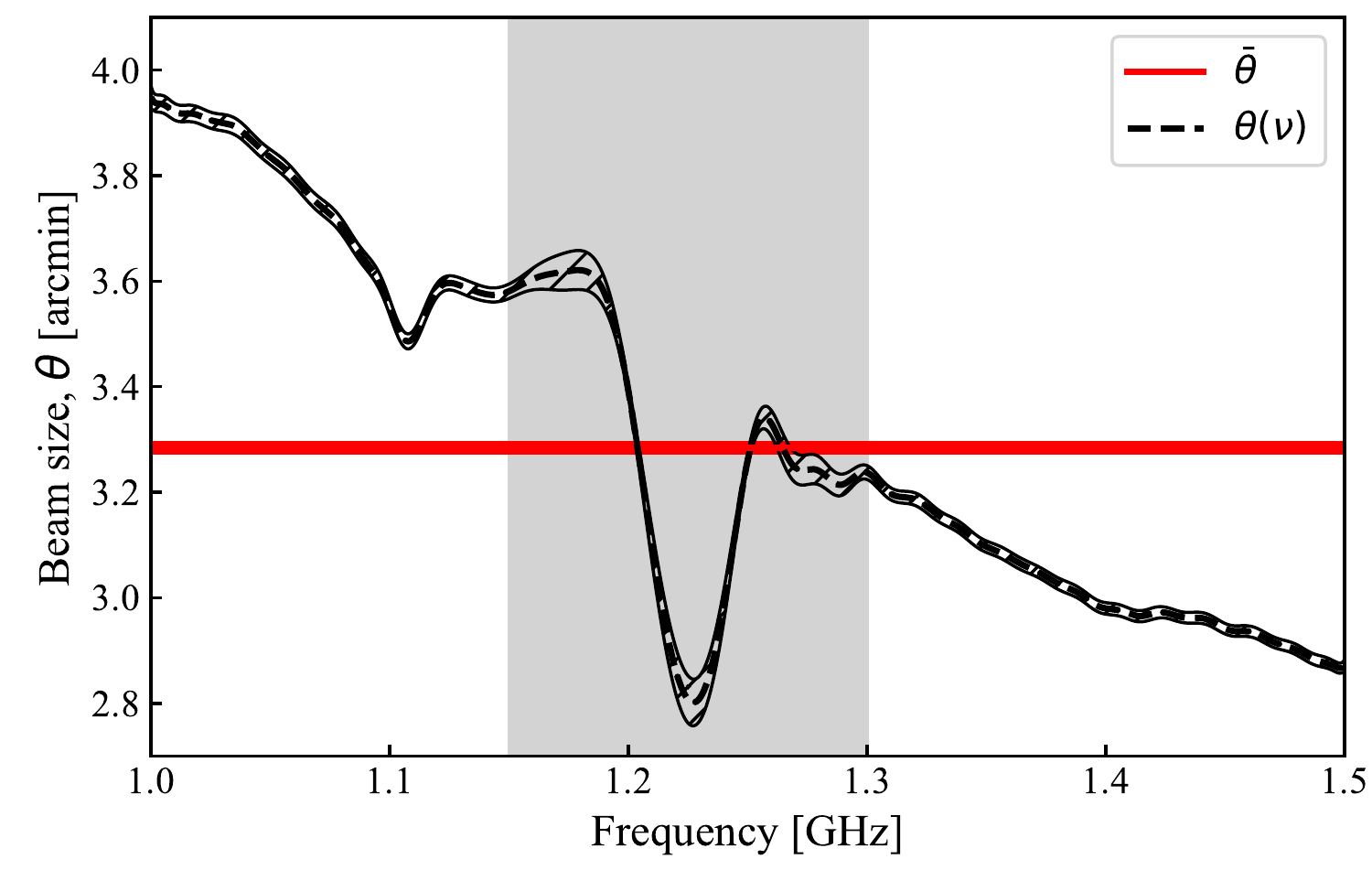}
                \caption{The variation of noise diode flux density (top left), SEFD (top right), pointing error (bottom left) and beam size (bottom right) as a function of frequency is shown in black for beam 1, polarisation XX$^*$. The hatched areas show the 1-$\sigma$ uncertainties. The grey area shows the frequency range from 1.15 to 1.3 GHz which is affected by RFI (GNSS satellites). The mean values, which exclude RFI-affected regions and edge channels, are shown in red.}\label{Fig_11}
            \end{center}
        \end{figure*}

        However, we also derive frequency-dependent parameters, $S_\mathrm{noise}(\nu)$, $S_\mathrm{sys}(\nu)$, $\sigma_\mathrm{pointing}(\nu)$, and $\theta(\nu)$ using the frequency-dependent flux density of calibrator, $S_\mathrm{calibrator}(\nu)$ derived in Section \ref{Sct_03_02_02}. Some uncertainty is introduced at frequencies affected by RFI, so stable results can only be derived using the baseline of raw spectrum provided in Section \ref{Sct_04_01}. In some cases (e.g. for beam size), the frequency-dependence is strong, as expected. In other cases (e.g. pointing error), there is virtually no frequency dependence. An example Gaussian fit to the beam response for polarization XX$^*$, beam 1 is shown in Figure~\ref{Fig_10}, and the derived instrumental parameters are shown in Figure~\ref{Fig_11}. Due to strong GNSS RFI the measurements between 1.15 and 1.30 GHz (the grey area Figure~\ref{Fig_11}) are less accurate.
        
        \subsubsection{Noise diode flux density}\label{Sct_05_02_01}
            
            In the upper left panel of Figure~\ref{Fig_11}, we show the flux density of noise diode. The mean flux density is $\bar{S}_\mathrm{noise}=0.72 \pm 0.08$ Jy, and the frequency-dependence,  $S_{\mathrm{noise}}(\nu)$ is small, rising between 1.08 and 1.12 GHz, and decreasing above 1.4 GHz. A similar variation is also found by \citet{2020RAA....20...64J}.
            
        \subsubsection{System equivalent flux density (SEFD)}\label{Sct_05_02_02}
            
            In the upper right panel of Figure~\ref{Fig_11}, we show the SEFD. The mean value is $\bar{S}_\mathrm{sys}=1.24 \pm 0.08$ Jy, but $S_\mathrm{sys}(\nu)$ increases by 15\% below 1.05 GHz, which is outside of the nominal frequency range. A bump is also seen between 1.08 and 1.12 GHz. Similar deviations are found in other beams and in the YY$^*$ polarisation. Since it also seems to correspond to a change in the SEFD and beamwidth, it is possibly an instrumental effect in the feed or the feed-LNA transition.
            
\begin{figure}
                \begin{center}
                    \includegraphics[width=\columnwidth]{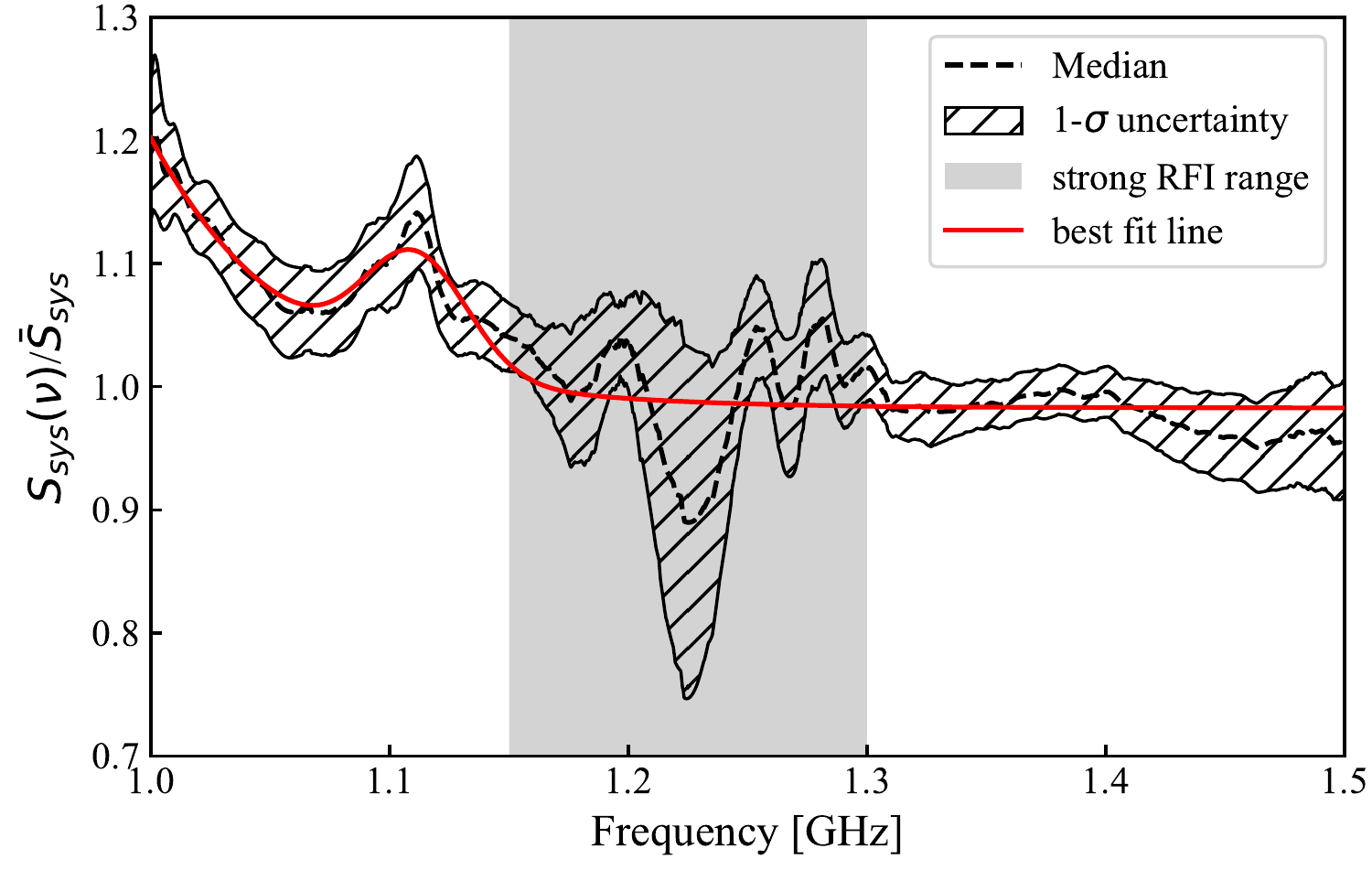}
                    \caption{The median normalised SEFD across all beams and polarisation channels is shown as a function of frequency (black dashed line), with corresponding  1-$\sigma$ uncertainty (hatched area). A rise is apparent at low frequency ($<1.15$ GHz) in all beams. We use Equation \ref{Equ_03} to fit the median, and derive the best fit line shown by the solid red line. Data affected by RFI (grey area) are excluded in the fit.}
                    \label{Fig_12}
                \end{center}
            \end{figure}

            In Figure~\ref{Fig_12}, we derive a median frequency-dependent correction factor for the mean SEFD correction by using data from both polarisation on all 19 beams. The correction factor is parameterised according to:
            \begin{equation}
                f_{\rm sys}(\nu) = A \times e^{-(\frac{\nu - \nu_0}{B})} + C \times e^{-(\frac{\nu - \nu_1}{D})^2} + E,
                \label{Equ_03}
            \end{equation}
            where the first term is the exponential rise at low frequency ($<1.05$ GHz), the second term is the Gaussian component at  1.1 GHz, and the third term is the deviation at high frequency ($>1.3$ GHz). We use the optimised $\chi^2$ method to derive the best fit parameters for the median line, excluding the RFI affected frequency range (1.15 -- 1.30 GHz). The derived best fit parameters are given in Table \ref{Tab_03} and the best fit line is shown by the red solid line. The uncertainty for each of the parameters is estimated using a jackknife method, i.e. removing one spectrum from 19 beams $\times$ 2 polarisation channels each time.
            
            \begin{table}
                \caption{The best fit parameters for the normalised frequency-dependence of the SEFD using Equation \ref{Equ_03}. The uncertainty is from the jackknife method.}
                \label{Tab_03}
                \centering
                \begin{tabular}{cc}
                    \hline\hline
                    Parameter & Value (jackknife uncertainty) \\
                    \hline
                    $\nu_0$ [GHz] & 0.98   (0.01)      \\
                    $\nu_1$ [GHz] & 1.1108 (0.0002) \\
                    $A$           & 0.28   (0.06)      \\
                    $B$           & 0.060  (0.002)  \\
                    $C$           & 0.094  (0.001)    \\
                    $D$           & 0.0306 (0.0006)  \\
                    $E$           & 0.9828 (0.0006)  \\
                    \hline\hline
                \end{tabular}
            \end{table}
        
\begin{figure}
                \begin{center}
                    \includegraphics[width=\columnwidth]{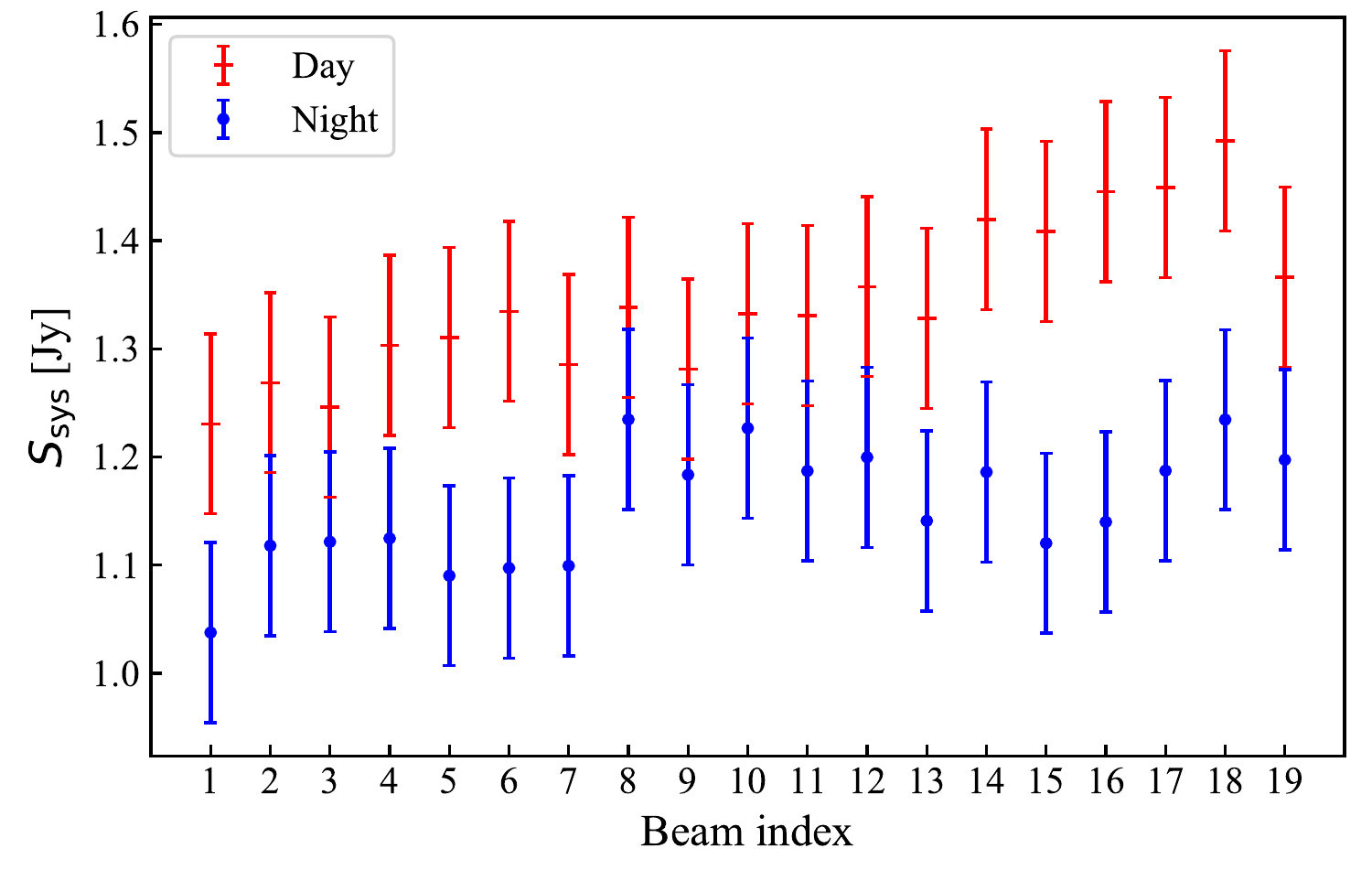}
                    \caption{The SEFD for all 19 beams is shown. The red pluses represent day time data, while the blue dots represent night time data. The beams in outer circles of the receiver have higher SEFD. The mean daytime SEFD is 1.34 Jy, and 1.15 Jy at night. }\label{Fig_13}
                \end{center}
            \end{figure}

            To evaluate the effects of solar interference on the SEFD, we compare the $\bar{S}_{\mathrm{sys}}$ of each beam for the day and night observations (see Figure~\ref{Fig_13}). We find values for $\bar{S}_{\mathrm{sys}}$ which are about 16\% higher in the day. The mean $\bar{S}_{\mathrm{sys}}$ values are 1.34 Jy and 1.15 Jy in the day and night, respectively.
            
            For daytime observations, we list the mean $\bar{S}_{\mathrm{sys}}$ and its uncertainty for different combinations of beams in Table~\ref{Tab_04}. The central beam (beam 1) has the lowest $\bar{S}_{\mathrm{sys}}$ value of 1.23 Jy. The inner circle of beams (beams 2--7) has a similar mean value ($S_{\rm sys} = 1.29$ Jy). The second circle (beams 9, 11, 13, 15, 17 and 19) has a significantly higher mean value ($\bar{S}_{\rm sys}=$1.36 Jy), and the final circle (beams 8, 10, 12, 14, 16 and 18) have the highest mean value ($\bar{S}_{\rm sys} = 1.40$ Jy). Table~\ref{Tab_04} shows that this is not due to a rise in system temperature $T_{\mathrm{sys}}$, which is fairly constant (here we adopt the gain at 1.35 GHz cited by \citet{2020RAA....20...64J} to convert SEFD to $T_{\mathrm{sys}}$). Instead, it is due to the classical drop in off-axis antenna efficiency for paraboloid antennas \citep{1965ITAP...13..660R}.
            
            \begin{table}
                \caption{The mean SEFD, system temperature and mean beam size for beams at different offsets from the optical axis (beam 1) for observations of 2019 Aug 25.}
                \label{Tab_04}
                \centering
                \begin{tabular}{lrrr}
                    \hline\hline
                    Position      & $\bar{S}_{\mathrm{sys}}$ ($\Delta \bar{S}_{\mathrm{sys}}$) & $\bar{T}_{\mathrm{sys}}$ ($\Delta \bar{T}_{\mathrm{sys}}$) & $\bar{\theta}$ ($\Delta\bar{\theta}$)  \\
                                           & [Jy]        & [K]        & [arcmin]    \\
                    \hline
                    Center                 & 1.23 (0.08) & 19.7 (1.4) & 3.26 (0.01) \\
                    1$^\mathrm{st}$ circle & 1.29 (0.08) & 19.6 (1.3) & 3.28 (0.01) \\
                    2$^\mathrm{nd}$ circle & 1.36 (0.08) & 19.7 (1.3) & 3.29 (0.02) \\
                    3$^\mathrm{rd}$ circle & 1.40 (0.08) & 20.0 (1.3) & 3.32 (0.03) \\
                    Any                    & 1.34 (0.08) & 19.8 (1.3) & 3.30 (0.02) \\
                    \hline\hline
                \end{tabular}
            \end{table}
        
        \subsubsection{Pointing error}\label{Sct_05_02_03}
        
\begin{figure}
                \begin{center}
                    \includegraphics[width=\columnwidth]{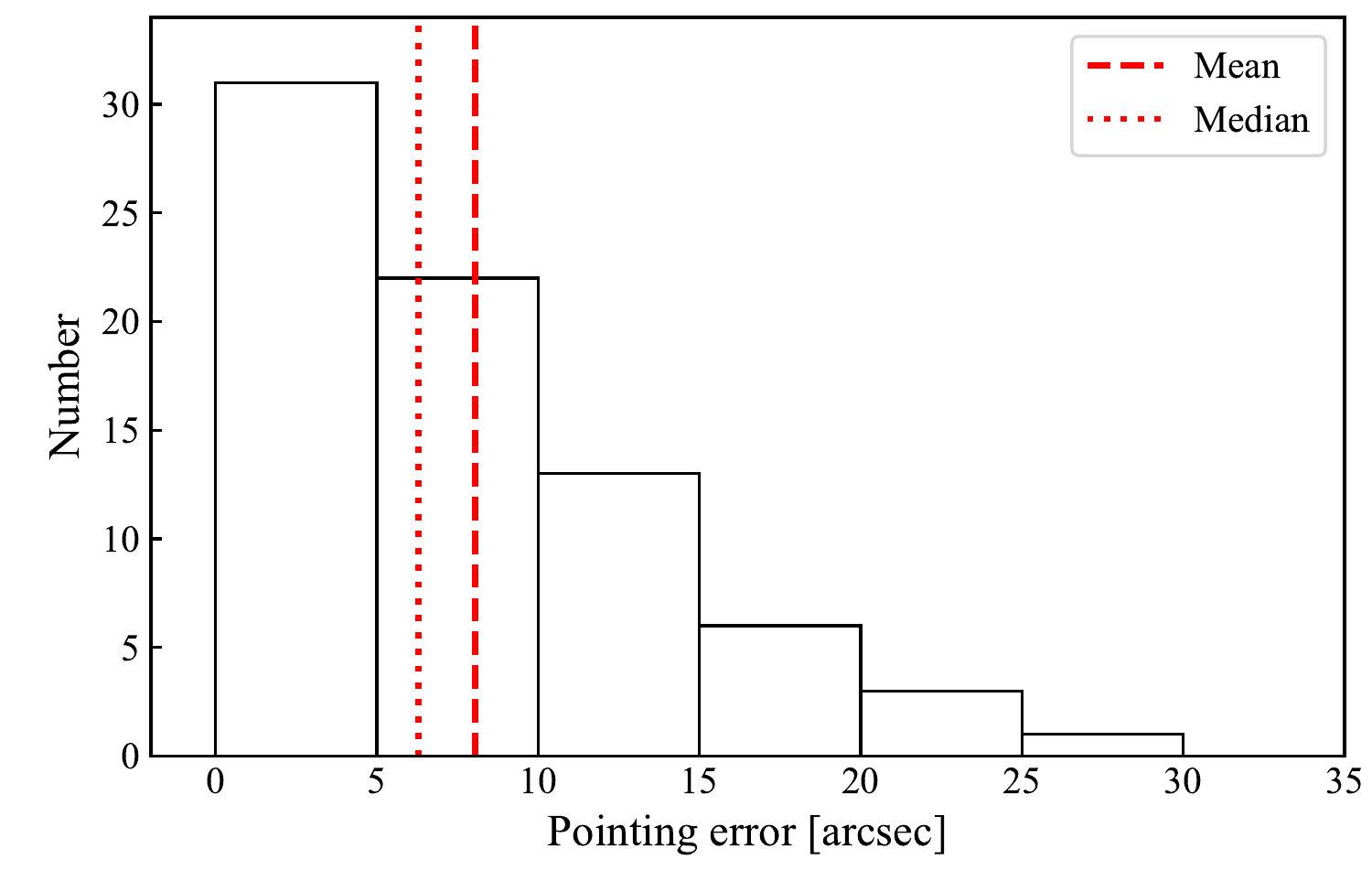}
                    \caption{A histogram of the pointing errors measured in both polarisations, all beams, and in both observing days. The bins have a width of 5 arcsec. The median and mean values are 6.30 and 8.05 arcsec, and are shown by red dotted line and red dashed line, respectively.}\label{Fig_14}
                \end{center}
            \end{figure}

            In the lower left panel of Figure \ref{Fig_11}, we show the measured pointing error. It clearly shows that $\sigma_\mathrm{pointing}(\nu)$ is nearly constant, and consistent with the mean value, $\bar{\sigma}_\mathrm{pointing}$. We also show the histogram of mean pointing errors across all beams and polarisations over two days of observations (76 measurements) in Figure \ref{Fig_14}. The mean and median errors are 6.3 arcsec and 8.1 arcsec, repsectively. Considering the 2.8 arcmin beam size of FAST at 1.5 GHz, the pointing accuracy is better than $\frac{1}{20}$ of the FAST beam.
            
        \subsubsection{Beam size}\label{Sct_05_02_04}
        
            The lower right panel of Figure \ref{Fig_11} shows the beam size measurements. The mean beam size $\bar{\theta}$, is 3.26 arcmin. The beam size for beams at different radii from the optical axis show a weakly increasing trend (see the final column in Table~\ref{Tab_04}). Our findings are consistent with the results from \citet{2020RAA....20...64J}.
            
\begin{figure}
                \begin{center}
                    \includegraphics[width=\columnwidth]{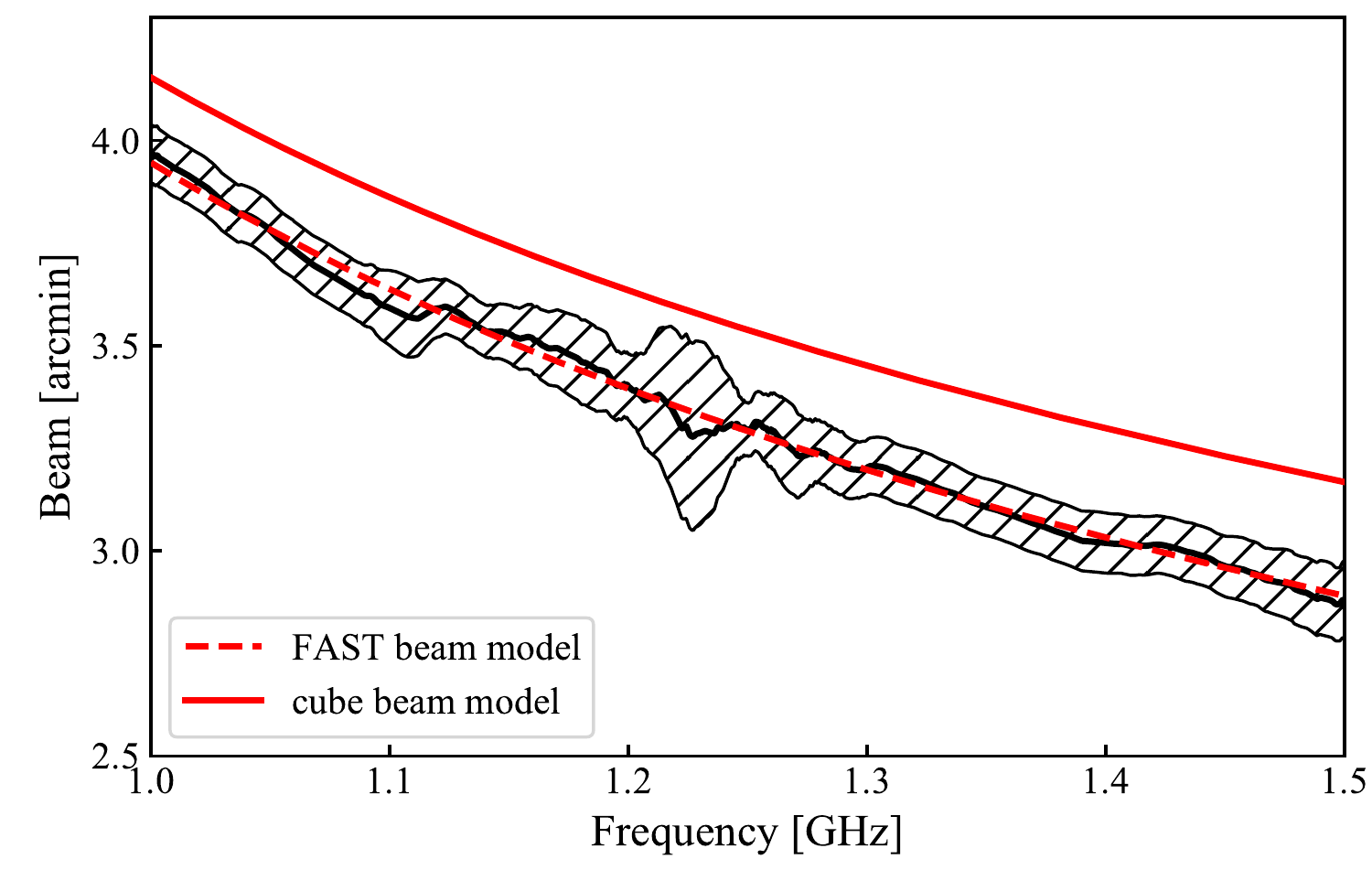}
                    \caption{The median FWHP beam size for all 19 beams is shown by the black solid line, with the corresponding 1-$\sigma$ uncertainty shown by the hatched area. The best fit line by using Equation \ref{Equ_04} is shown by the red dashed line. The beam size in our final cube (post-gridding, described in Section \ref{Sct_06_04}) is shown by the red solid line.
                    }
                    \label{Fig_15}
                \end{center}
            \end{figure}

            As expected, the beam size $\theta(\nu)$, decreases with frequency. Across all beams and both polarisations, the median frequency-dependent beam size and its 1-$\sigma$ uncertainty is shown in Figure~\ref{Fig_15}.
            The beam size is dependent on wavelength and telescope diameter. For an unblocked circular aperture with uniform illumination, the FWHP beam size is $\theta = 1.02 \lambda/D$ in radians \citep{2013tra..book.....W},  where $\lambda$ is the wavelength and $D$ is the illuminated diameter. Normal radio telescopes with tapered illumination patterns will typically have beam sizes which are larger by 20\%, but this depends on the taper \citep{2013tra..book.....W} and the offset from the optical axis.  For the similar prime-focus Parkes multibeam system \citep{1996PASA...13..243S}, the average FWHP beam size is $\theta = 1.225 \lambda/D$ in radians. Hence, for the multiple beams on FAST, we parameterise the beam size as follows:
            \begin{equation}
                \theta = 1.225 \frac{\lambda}{A + B\left( \frac{\lambda-0.21}{0.21}\right) + C\left(\frac{\lambda-0.21}{0.21}\right)^2},
                \label{Equ_04}
            \end{equation}
            where $A$, $B$ and $C$ are functions of $\lambda$. We use an optimised $\chi^2$ method to find the best fit (see Figure \ref{Fig_15}). The best fit parameters and the parameters for all the 19 beams are listed in Table \ref{Tab_05} and Table~\ref{Tab_06}, respectively.
        
            \begin{table}
                \caption{Parameters of Equation \ref{Equ_04} for the beam size of FAST, and beam size in our final cube.}
                \label{Tab_05}
                \centering
                \begin{tabular}{crrr}
                    \hline\hline
                    \multirow{2}{*}{} & \multicolumn{3}{c}{Parameters}\\
                    \cmidrule{2-4}
                    & A [m] & B [m] & C [m]\\
                    \hline
                    Model      & 295.69 (0.05) & 89.42 (0.67) & -76.47 (1.70) \\
                    Final cube & 271.27 (0.04) & 111.44 (0.53) & -81.83 (1.37) \\
                    \hline\hline
                \end{tabular}
            \end{table}

\section{Data Reduction}\label{Sct_06}
        
    \subsection{System quivalent flux density}\label{Sct_06_01}
        
        For target field flux density calibration, we use the flux density of noise diode as calibrated by the above calibrator observations. We first remove channels affected by the known RFI and channels outside the FAST nominal bandpass ($<1.05$ or $>1.45$ GHz) (for calibration purposes only). Subsequently, we calculate the mean power for all the spectra. We derive the CAL step by subtracting the CAL off spectrum from the CAL on spectrum that it follows. The value for $\bar{S}_\mathrm{noise}$ derived in Section \ref{Sct_05_02} is used as the equivalent flux density of the CAL step. The derived value for $\bar{S}_{\mathrm{sys}}$ is then used for the subsequent 59 spectra. The frequency-dependent scale factor derived in Section \ref{Sct_05_02} is applied after flagging.
        
    \subsection{Bandpass removal}\label{Sct_06_02}
        
\begin{figure*}
            \begin{center}
            
                \includegraphics[width=0.45\columnwidth]{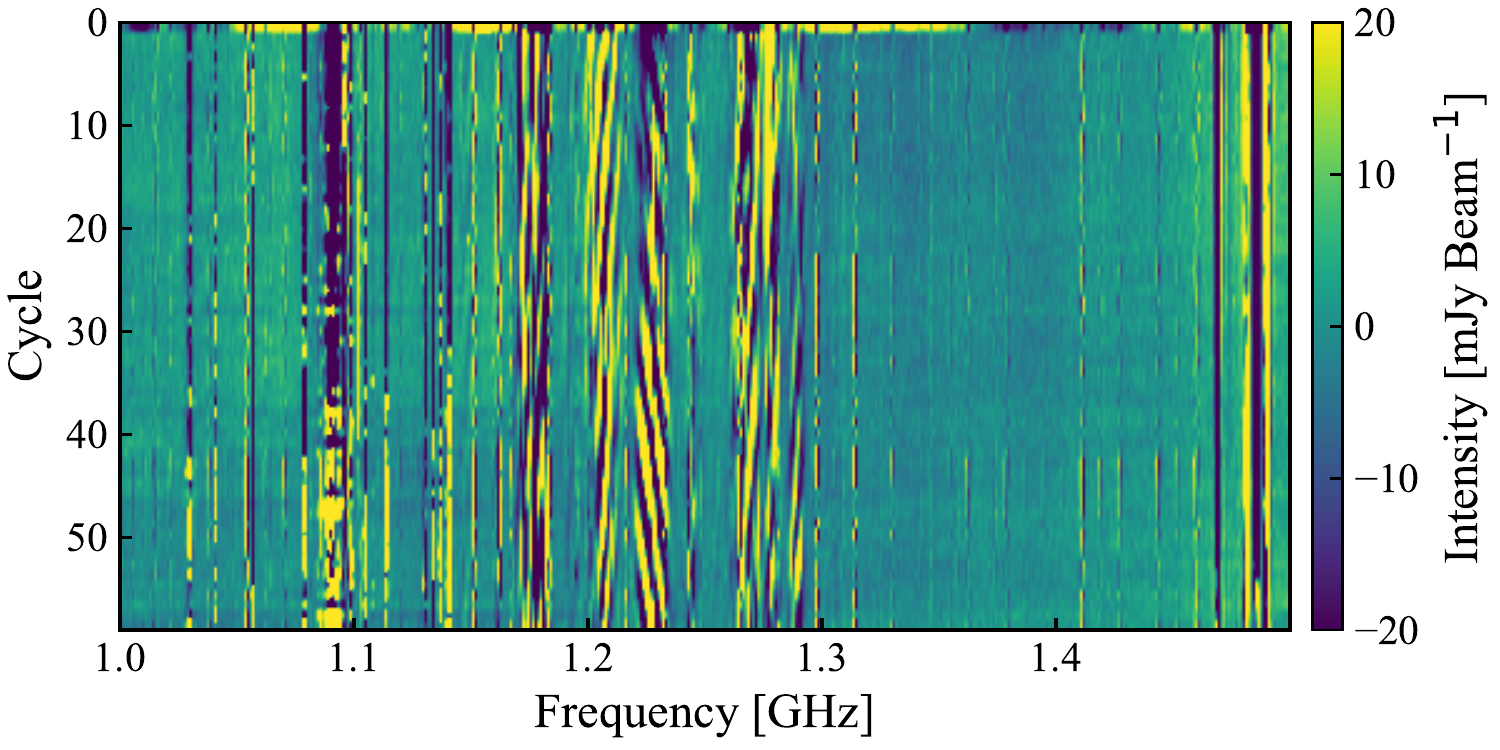}
                \includegraphics[width=0.45\columnwidth]{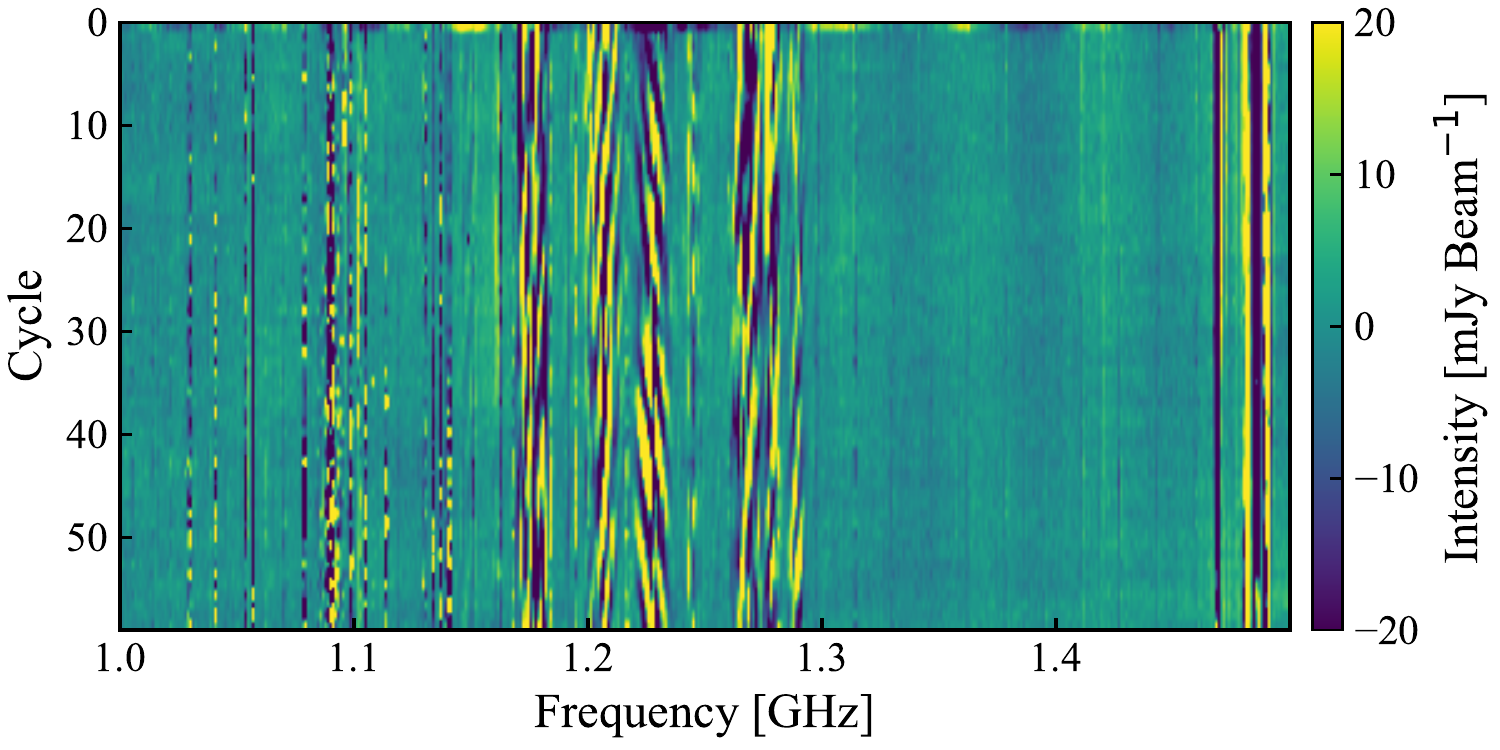}
                \includegraphics[width=0.45\columnwidth]{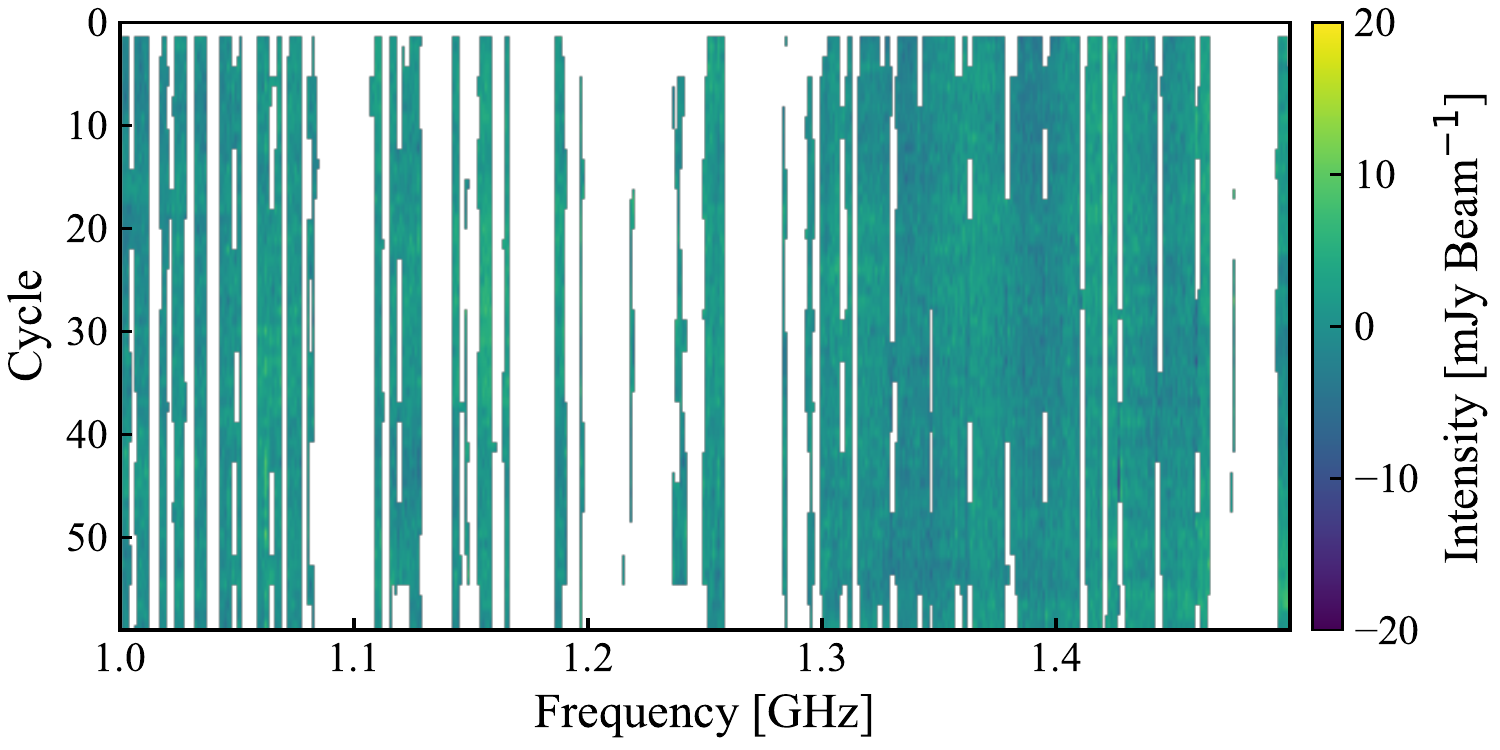}
                \includegraphics[width=0.45\columnwidth]{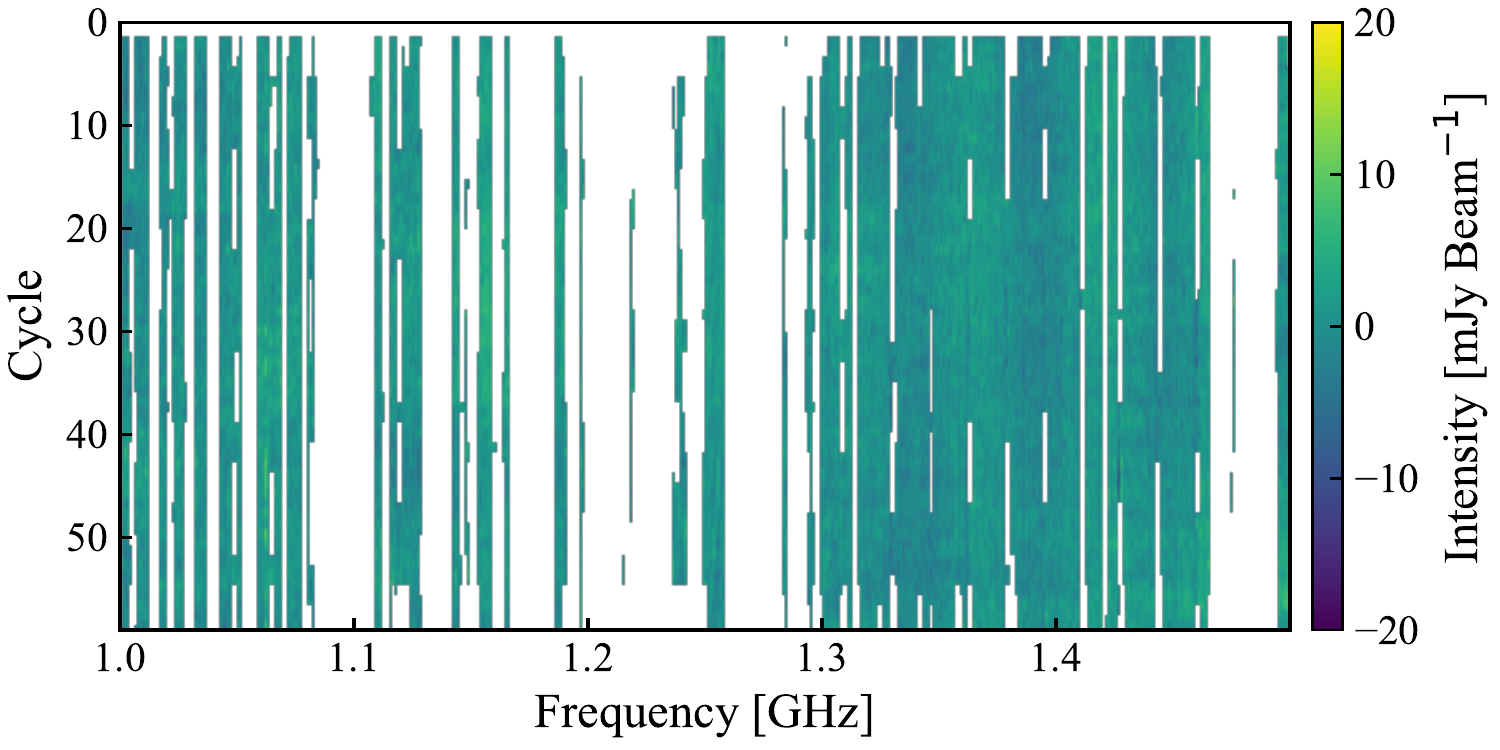}
                
                \caption{Upper left panel: calibrated data for polarisation XX$^*$, beam 1 in the frequency-time domain. Upper right panel: constant RFI is suppressed by removing the median spectrum across the 60 integration cycles. Lower left: data flagged as RFI is removed. Lower right panel: final frequency-dependent  intensity calibration is applied.
                }\label{Fig_16}
                
            \end{center}
        \end{figure*}

        We follow the robust bandpass calibration procedure as used for HIPASS \citep{2001MNRAS.322..486B}. In order to calibrate a raw spectrum $P_i(\nu)$ according to its measured SEFD  $S_i$, we select neighbouring spectra as a reference spectrum set, $C_r$. The reference SEFD $S_{\mathrm{sys},r}$, and the reference spectrum $P_r(\nu)$, are computed from the median of the spectra in $C_r$. Subsequently, the raw spectrum is flattened using the reference spectrum and scaled to units of $S_{\mathrm{sys},r}$. The system temperature contribution, including sky continuum, is then removed by subtracting $S_{sys, i}$. The procedure is summarised by:
        \begin{equation}
            S_i(\nu) = S_{\mathrm{sys}, r} \times \frac{P_i(\nu)}{P_r(\nu)} - S_{\mathrm{sys}, i},
            \label{Equ_05}
        \end{equation}
        where $S_i(\nu)$ is the calibrated spectrum in Jy, $S_{\mathrm{sys}, i}$ and $S_{\mathrm{sys}, r}$ are in Jy, and $P_i(\nu)$ and $P_r(\nu)$ are in (arbitrary) units of power. For each beam, we include the nearest 260 spectra in the same beam as the reference spectrum set, giving spatial coverage of about two scan lines. As explained in \citet{2014PASA...31....7C}, the form of Eq.~\ref{Equ_05} flattens the bandpass, converts the flux scale to Jy and removes continuum to zeroth-order. Due to residual spectral artefacts (spectral index and ripple) in strong continuum sources ($>10$ mJy), spectra within 3 arcmin of these sources are excluded in the reference spectrum set. To reduce sidelobes caused by any strong HI emission, we use a 5-channel Hanning window (0.25, 0.75, 1, 0.75, 0.25) to convolve with the calibrated spectra. Therefore, the frequency resolution is lowered to 22.9 kHz. This procedure is performed for each polarisation channel and for all 19 beams. We use a cubic polynomial function to fit the calibrated spectra to further flatten the baseline. Figure \ref{Fig_16} shows 60 calibrated spectra for the XX$^*$ polarisation of beam 1. The residual compressor RFI, particularly the vertical strips between 1.31 and 1.41 GHz, are still visible (upper left panel). The use of the median spectrum for all 19 beams suppresses the compressor RFI (upper right panel) but also the extended Galactic \HI\ emission\footnote{Raw data are archived, so re-processing is possible for future Galactic \HI\ studies.}.
        
    \subsection{RFI removal}\label{Sct_06_03}
        
        To remove residual RFI from all target spectra, we adopt the following flagging steps: 
    
        \begin{enumerate}
            
            \item Pixels without extreme values (not deviating more than $\pm$0.1 Jy) are used for deriving noise. The theoretical sensitivity for the given frequency and time resolution is 15.3 mJy, assuming a SEFD of 1.3 Jy (corresponding to a gain of 15 K Jy$^{-1}$ and a system temperature of 20 K).
            
            \item Channels with high noise levels, $\sigma_c$ > $1.2\sigma_g$ are noted as bad, where $\sigma_c$ is computed as 1.4826$\times$MAD for each channel in all scans in one dimension (either RA or Dec) over one day, and $\sigma_g$ is the global average calculated in the same way over all channels.
                
            \item Pixels in bad channels with high noise levels in the time-frequency domain are flagged, i.e. $\sigma_p$ > $2\sigma_g$, where $\sigma_p$ is defined as the RMS in the surrounding $5 \times 5$ box. This step also detects regions where the mean flux density deviates from zero, so bright galaxies in RFI-affected channels may be eliminated.
                
            \item Any flagged pixel is grown into a flagged $5 \times 5$ pixel region.
                
            \item The spectra take with CAL on, which have higher $S_{\mathrm{sys}}$, are also flagged. 
                
        \end{enumerate}
    
        The third panel of Figure \ref{Fig_16} shows the effect of removing the flagged pixels. The flagged data accounts for 25.6\%. of the calibrated data, which is smaller than the 41.7\% of flagged RFI data in the raw spectral data (Figure \ref{Fig_05}). This noise-based flagging method therefore recovers 15\% of the data that would otherwise have been flagged. This is partly helped by the compressor RFI template. As already noted, the frequency-dependent flux density correction factor, $f_\mathrm{sys}(\nu)$ is applied to the data after flagging, and also shown in Figure~\ref{Fig_16}.

    \subsection{Gridding}\label{Sct_06_04}
    
        We employ the robust gridding algorithm developed for HIPASS to create our final FUDS data cube. The final cube has a size of 1\degr$\times$ 1\degr and a pixel size of 1 arcmin$\times$1 arcmin. The frequency spacing in the final data cube is the same as the raw spectra, 7.63 kHz, but the resolution after Hanning smoothing is 22.9 kHz. For each pixel, we use a radius of 1.3 arcmin as a cutoff to select the contributing spectra. Considering the variation of beam size with frequency, up to 40\% across the whole bandwidth, the median frequency-dependent beam size derived in Section \ref{Sct_05_02} is used for all beams. The spectrum for a given position in the cube is then derived as follows:
        \begin{equation}
            S(\nu)=\frac{\mathrm{median}(S_r(\nu))}{\mathrm{median}(w_r(\nu))},
            \label{Equ_06}
        \end{equation}
        where $S_r(\nu)$ are the spectra within 1.3 arcmin and $w_r(\nu)$ are their corresponding Gaussian weights.
        
        Assuming all the 19 beams have same analytic beam size, $\theta(\nu)$, derived in Section \ref{Sct_05_02}, we perform the same gridding procedure above to derive the beam size in our final cube by using a Monte Carlo (MC) method. We then use Equation~\ref{Equ_04} to fit the gridded beam. The best fit line is shown in Figure~\ref{Fig_15}, and the best fit parameters of Equation~\ref{Equ_04} are given in Table~\ref{Tab_05}. This beam size is required when deriving precise flux densities for extended source in the gridded cubes.
        
    \subsection{Strong continuum sources}\label{Sct_06_05}
        
        We employ the method of \citet{2001MNRAS.322..486B} to remove residual emission near strong continuum sources. We use the spectra toward continuum sources with flux densities stronger than 10 mJy to generate a template spectrum, given by: 
        \begin{equation}
            \mathbf{S}_T(\nu) = \Sigma_i w_i S_i(\nu),
            \label{Equ_07}
        \end{equation}
        where $S_i(\nu)$ is the central spectrum of $i$th continuum source, and $w_i$ is the weight. The spectrum associated with the strongest continuum source has a weight set to unity. The weight of other sources is determined by the slope of the linear regression between two spectra $S_i(\nu)$ and $S_1(\nu)$ in the frequency ranges 1.015 -- 1.137 GHz, 1.30 -- 1.416 GHz and 1.423 -- 1.463 GHz. For each spectrum in the final data cube, we perform a linear regression with the template in the same manner to derive the weight $w$, then subtract $w\mathbf{S}_T(\nu)$ from the spectrum.

\section{Confusion}\label{Sct_07}
    
    Confusion happens when multiple galaxies cannot be distinguished in both space and velocity due to the limited resolution of a telescope. There are two main effects on a survey: 1) an underestimate of the number of observable galaxies, and 2) an overestimate of the \HI\ mass due to the contribution of unresolved galaxies.  \citet{2016MNRAS.455.1574J} shows an example of the increase in apparent HI mass with increasing beam size. Based on the estimates from  stacking, they predict a confused mass of $\sim 4 \times 10^{9}~h_{70}^{-2}{\rm M_\odot}$ within a FAST-like beam size (3 arcmin) for a galaxy with velocity width of 600~km~s$^{-1}$ at $z=0.1$. \citet{2019SCPMA..6259506Z} employed the same method to estimate the confused mass within the FAST beam for the CRAFTS survey, and found a similar confused mass of $\sim 2\times 10^{9}~h_{70}^{-2}{\rm M_\odot}$ at a similar redshift.
    
    To investigate the confusion in FUDS, we perform a simulation following the method described in \citet{2015MNRAS.449.1856J}. In the simulation, we assume a FUDS sensitivity of $50~\mu$Jy~beam$^{-1}$. First, we sample the galaxies based on the \HI\ mass derived from the HIPASS HIMF and the velocity widths sampled from the conditional velocity function \citep{2010ApJ...723.1359M}. We limit the galaxy masses between 10$^6$ and 10$^{12}~h_{70}^{-2} \rm M_\odot$ and the velocity widths between 15 and 10$^3$~km~s$^{-1}$. The correlation function is used to distribute the mock galaxies in 50$^3~h_{70}^{-1} {\rm Mpc}$ co-moving volumes at each redshift, starting with one galaxy located at each centre. The beam size in our final cube (see Section \ref{Sct_05_02_04}) is used as the confusion limit for angular separation. The distance along the line-of-sight is converted to a velocity separation using the Hubble-Lemaitre law. Subsequently, we perform the simulation 10$^4$ times to calculate mean values, independent of cosmic variance.
    
\begin{figure}
        \begin{center}
            
            \includegraphics[width=\columnwidth]{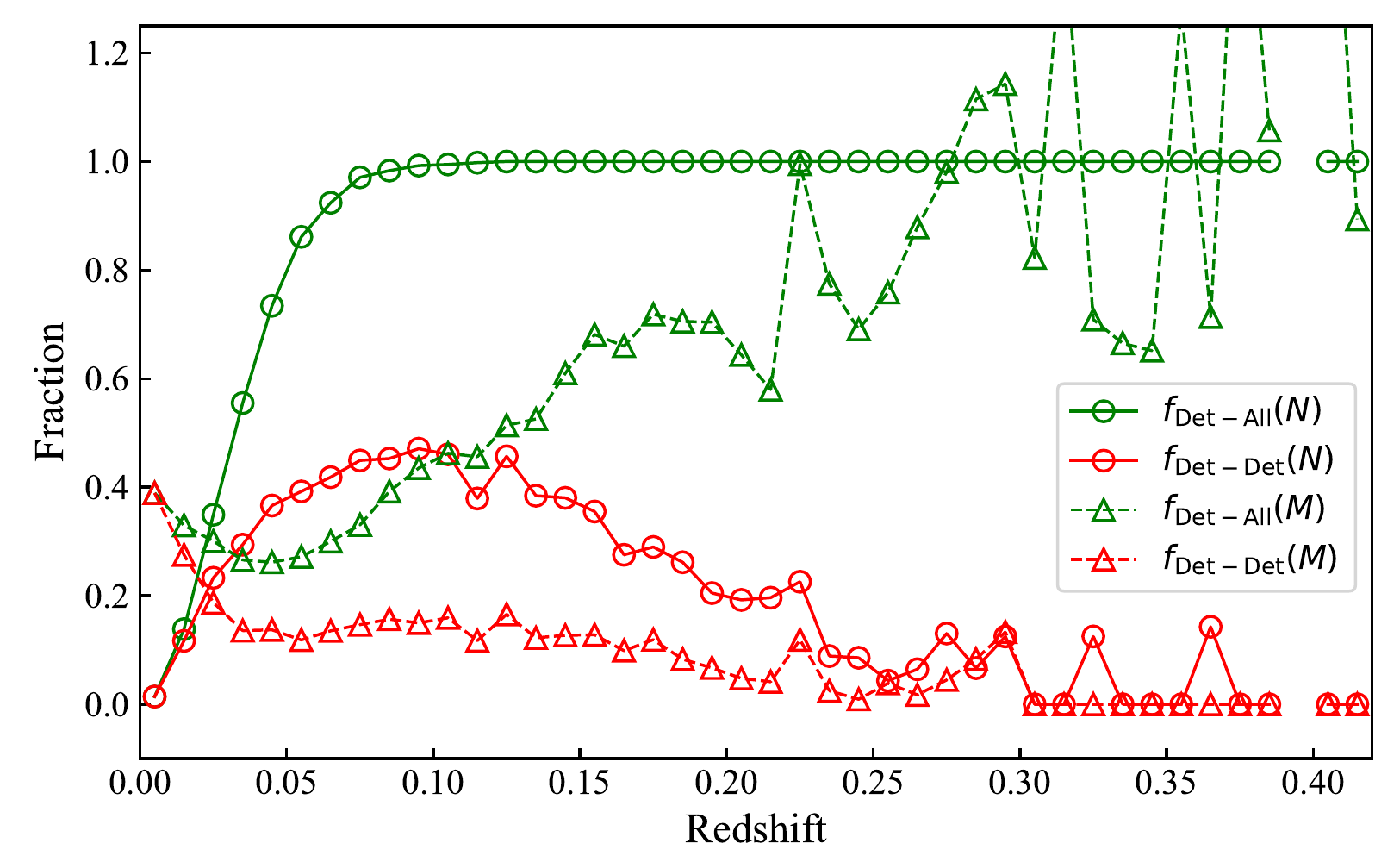}
            \caption{The green circles represent the confusion number rate for detected galaxies blended with any other galaxies ($f_{\rm Det-All}(N)$), while the red circles represent the confusion rate for detected galaxies blended with any other detected galaxies ($f_{\rm Det-Det}(N)$). The green triangles show the fraction of confused mass from all the confused galaxies for detected galaxies ($f_{\rm Det-All}(M)$). And the red triangles show the fraction of confused mass from all the detected confused galaxies for detected galaxies ($f_{\rm Det-Det}(M)$).}\label{Fig_17}
                
        \end{center}
    \end{figure}

    Here, we inspect the impact of confusion only for the detected galaxies. We define $f_{\rm Det-All}(N)$ and $f_{\rm Det-Det}(N)$ as the fraction of detected galaxies which are confused with other simulated galaxies and the fraction confused with only other detected galaxies, respectively. The definitions are:
    \begin{equation}
        f_{\rm Det-All/Det}(N) = \frac{N_{\rm Det-All/Det}}{N_{\rm Det}},
    \end{equation}
    where $N_{\rm Det-All}$ is the number of detected galaxies confused with any other galaxy, $N_{\rm Det-Det}$ is the number of detected galaxies confused with any other detected galaxy, and $N_{\rm Det}$ is the number of detected galaxies. The latter ratio, $f_{\rm Det-Det}(N)$, permits an estimate of the impact of confusion on the number of detected galaxies in FUDS.
    In Figure \ref{Fig_17}, we use green and red circles to show predicted values for $f_{\rm Det-All}(N)$ and $f_{\rm Det-Det}(N)$ in each redshift bin, respectively. $f_{\rm Det-All}(N)$ increases quickly to 100\% at $z \sim 0.1$. However, $f_{\rm Det-Det}(N)$ is much lower. At low redshift, it is negligible, but reaches a peak value of 40\% at $z=0.1$. For $z>0.1$, it drops sharply with $f_{\rm Det-Det}(N)$ being below 15\% at $z>0.25$. Therefore, at all redshifts, confusion will reduce the number of detections by less than 20\% (40/2)\%. Confusion will also increase the apparent \HI\ mass by merging with confused galaxies. Considering the galaxies below the detection limit, we find that the average total confused mass in a beam for our simulated central galaxies is $\sim 2\times 10^{9}~h_{70}^{-2}{\rm M_\odot}$ at $z=0.1$, and $10^{10}~h_{70}^{-2}{\rm M_\odot}$ at $z=0.4$. This is similar to the results of \citet{2019SCPMA..6259506Z}, and slightly lower than the values predicted in \citet{2016MNRAS.455.1574J}, probably due to taking into account the velocity distribution.
    
    In FUDS, we are interested in the corresponding ratio of the confused mass to the mass of the largest galaxy in each beam.
    We introduce $f_{\rm Det-All}(M)$ and $f_{\rm Det-Det}(M)$ to represent 
    the fraction of mass in the largest detected galaxy which is confused with mass in other simulated galaxies and the fraction confused with only mass in other detected galaxies, respectively. The definitions are:
    \begin{equation}
        f_{\rm Det-All/Det}(M) = \frac{M_{\rm Det-All/Det}}{M_{\rm Det}},
    \end{equation}
    where $M_{\rm Det-All}$ is the mass in all the confused galaxies except the largest one, $M_{\rm Det-Det}$ is the mass in detected confused galaxies except the largest one, and $M_{\rm Det}$ is the mass of the largest detected confused galaxy. The two parameters show us the deviation of measured \HI\ mass caused by confusion. In Figure \ref{Fig_17}, we use green and red triangles to show the average values $\bar{f}_{\rm Det-All}(M)$ and $\bar{f}_{\rm Det-Det}(M)$ across all our simulations as a function of redshift. We find that $\bar{f}_{\rm Det-All}(M)$ slowly increases with redshift, and reaches about 70\% at $z \sim 0.15$. The fluctuation beyond $z \sim 0.25$ is the result of less samples ($<20$) in each of these redshift bins. The mean value $\bar{f}_{\rm Det-All}(M)$ is about 95\% at high redshifts. However, confusion from undetected galaxies tends to elevate the spectral baseline, so does not necessarily result in an increase in the measured \HI\ mass. $\bar{f}_{\rm Det-Det}(M)$ may therefore be a more appropriate statistic to indicate the impact of confusion on measured \HI\ mass. $\bar{f}_{\rm Det-Det}(M)$ is about 15\% at low redshift ($0.04<z<0.15$), and almost zero at higher redshift ($z>0.25$). Note that both $\bar{f}_{\rm Det-All}(M)$ and $\bar{f}_{\rm Det-Det}(M)$ become large ($\sim$30\%) at very low redshift ($z<0.03$). That is caused by the dramatic increase in the numbers of low-mass galaxies which are detected. However, the cosmic volume at low redshift in FUDS is small. In reality, the contribution from confused mass will probably lie somewhere between $\bar{f}_{\rm Det-All}(M)$ and $\bar{f}_{\rm Det-Det}(M)$.
    
\begin{figure}
        \begin{center}
            
            \includegraphics[width=\columnwidth]{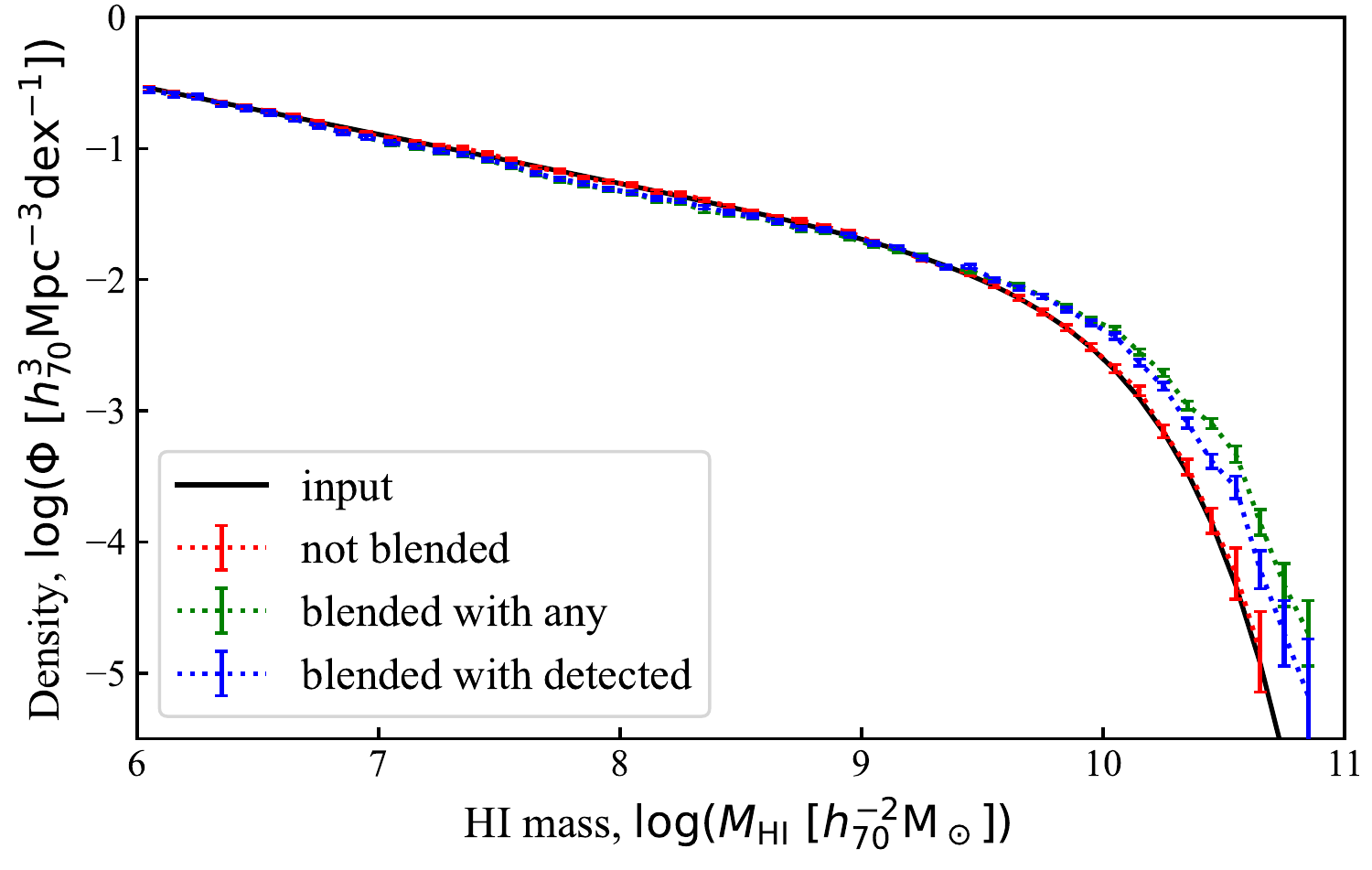}
            \caption{Three HI mass functions (HIMFs) are constructed from our simulated catalogue: 1) with no confusion (red); 2) confusing mass from all the confused galaxies (green); 3) confusing mass from the detected confused galaxies (red). The black line is the input HIMF.
            }\label{Fig_18}
        \end{center}
    \end{figure}

    In order to understand the impact of confusion on the measured HIMF, we recover the HIMF from the simulated catalogue using the $\Sigma 1/V_{\rm max}$ method \citep{1968ApJ...151..393S}. As  mentioned above, the contribution of confusing mass should be between $\bar{f}_{\rm Det-All}(M)$ and $\bar{f}_{\rm Det-Det}(M)$. Hence, we calculate the HIMF in three cases: 1) no confusion; 2) confusing mass from all the confused galaxies; 3) confusing mass only from the detected confused galaxies. The first is a reality check of the $\Sigma 1/V_{\rm max}$ method. The second and third cases give upper and lower limits to the contribution of confused mass. For simplicity, we assume that the confusion only increases flux, without changing the velocity width. Thanks to the large simulated sample, the uncertainties in each mass bin are low. The recovered HIMFs are shown in Figure \ref{Fig_18}. We use the Schechter function \citep{1976ApJ...203..297S} to parameterize the HIMFs. In the first case, the HIMF can be recovered with only a small deviation ($\Delta \alpha \sim 0.002$, $\Delta \log M^* \sim 0.003$ dex). When we take into account the confusing mass, the characteristic mass is overestimated by 0.20 - 0.29 dex. However, the low-mass end slope can is recovered well, and is only steepened by $-0.01$ to $-0.001$. The deviation in $\alpha$ is much smaller than 1-$\sigma$ uncertainty expected from a single FUDS field, while the deviation in $M^*$ is close to 2-$\sigma$. A similar trend was noted by \citet{2015MNRAS.449.1856J}. Since the cosmic \HI\ density  $\Omega_{\rm \HI}$ is often estimated by integrating the \HI\ mass-weighted HIMF, confusion can potentially impact $\Omega_{\rm \HI}$ calculations. In our simulation, $\Omega_{\rm \HI}$ is recovered with an accuracy of 7\%. However, $\Omega_{\rm \HI}$ is overestimated by 23 - 33\% when considering confusion. In practice, the effect of confusion will be more complicated than modelled by this simulation due to the effects of cosmic variance, clustering, non-zero peculiar velocities and environmental effects. The effect on `stacking' studies is also much more dramatic than for directly detected galaxies \citep{2013MNRAS.433.1398D}, although the availability of photometric and spectroscopic redshifts can partly mitigate its effects by allowing for field-by-field estimates of confusion. Further and more detailed consideration of confusion will be presented in a future paper.
    
\section{Preliminary results for the FUDS0 field}\label{Sct_08}
    
    Observations of the FUDS0 field have been concluded in our pilot survey. The on-source integration time is about 90 hrs. The calibration, data reduction methods described above are employed to process the data. The rms for the final data cube is $\sim 50~\mu$Jy in the RFI-free regions and at the centre of the field. We detect 128 galaxies in \HI\ with redshifts in the range of $0<z<0.4$ and \HI\ masses in the range of $6.8<\log(\frac{\rm M_{\HI}}{h_{70}^{-2} M_{\odot}})<11.0$.
    
\begin{figure}
        \begin{center}
            
            \includegraphics[width=\columnwidth]{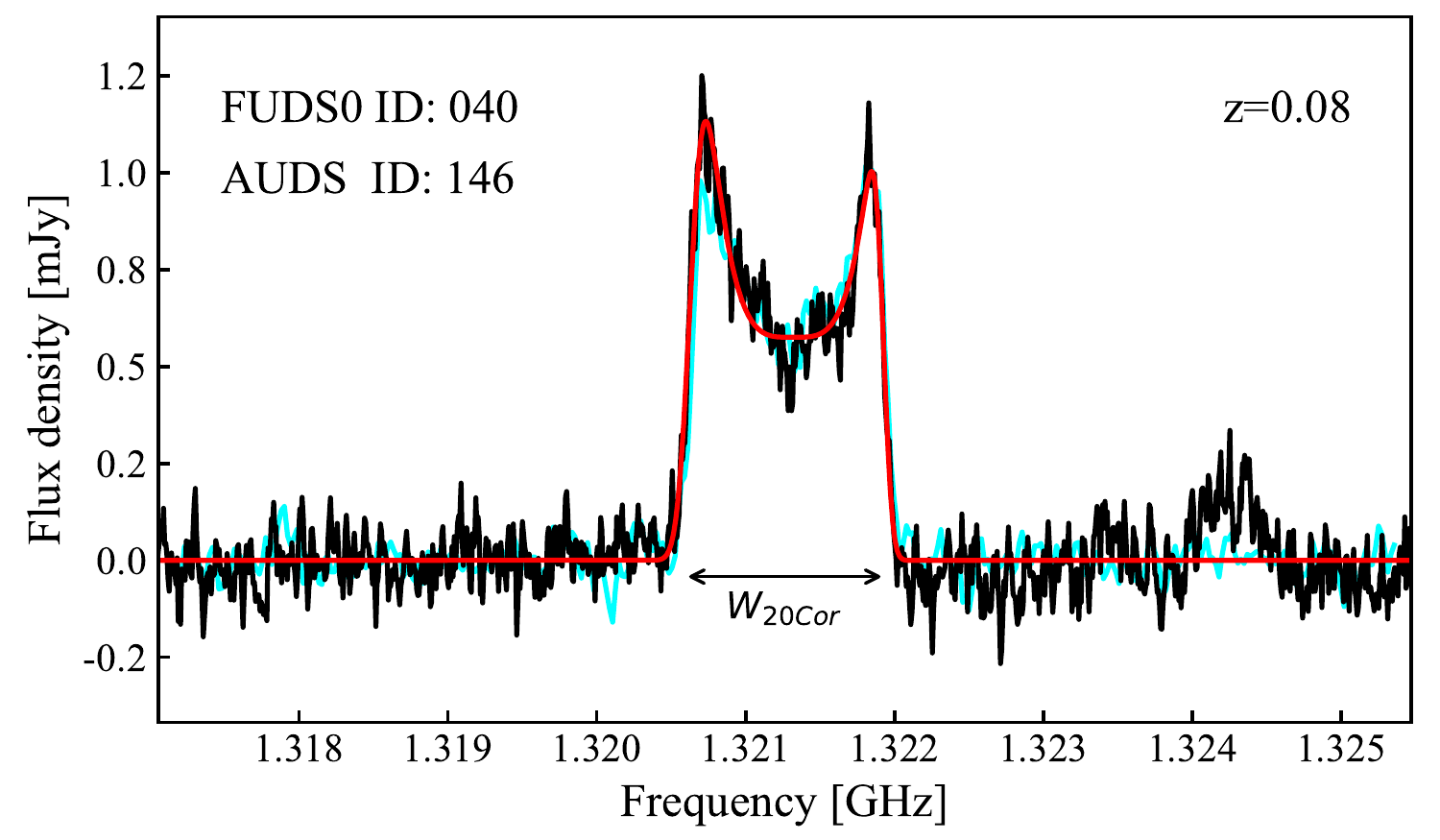}
            \includegraphics[width=\columnwidth]{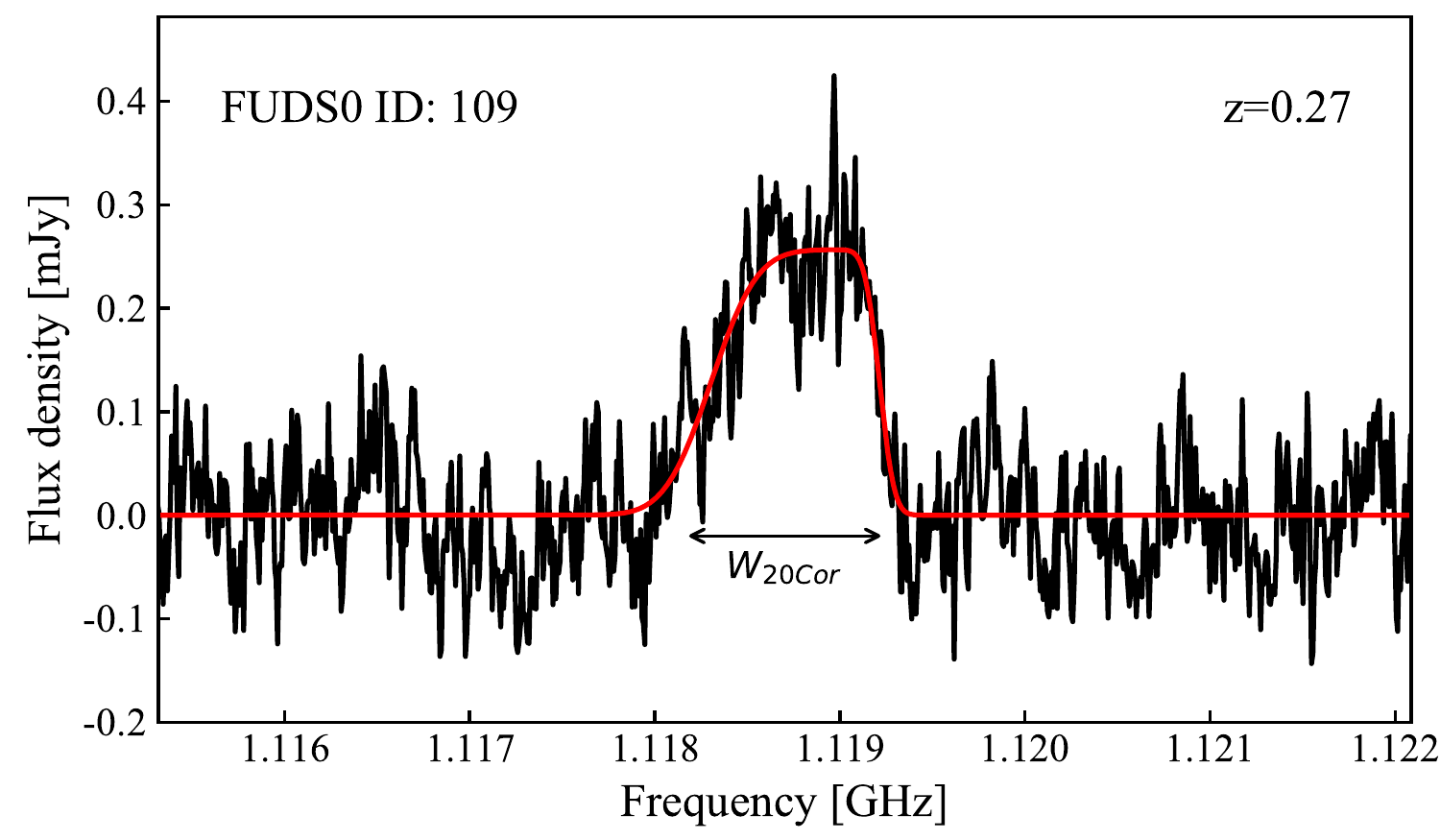}
            \caption{Example spectra of two galaxies detected in the FUDS0 field (black lines). The ID and redshift are given in upper left and right corners, respectively. A Busy function is used to fit the spectra (red lines). The galaxy the upper panel is also detected in AUDS. The AUDS spectrum is overlaid (cyan line), and its AUDS100 ID is shown in the upper left corner. In the lower panel, we show an example spectrum of a galaxy at $z=0.27$, which is beyond the AUDS redshift limit of 0.16.}\label{Fig_19}
                
        \end{center}
    \end{figure}

    In Figure \ref{Fig_19}, we show the spectra of two of the detected galaxies, with a Busy Function fit \citep{2014MNRAS.438.1176W} overlaid. Since the FUDS0 field partially overlaps with GAL2577 field in the Arecibo Ultra-Deep Survey (AUDS) \citep{2021MNRAS.501.4550X}, 33 galaxies with $z<0.16$ (the AUDS limit) are detected in both surveys. A comparison of integrated flux and line width was performed between the two surveys. And we did not find any systematic errors. FUDS 040, in the upper panel is one of these. The AUDS spectrum is overlaid. The spectra from both surveys are consistent, which provides confidence in our calibration and data reduction methods. FUDS 109 in the lower panel lies at $z=0.27$, which is beyond the redshift limit of AUDS. Detailed results from the FUDS0 field will be presented in a separate paper.
    
\section{Summary}\label{Sct_09}

    We introduce a new deep extragalactic \HI\ survey on the FAST telescope, the FAST Ultra-Deep Survey (FUDS). The high sensitivity of FAST and wide frequency range of its multibeam receiver enables studies of high redshift galaxies, faint nearby objects, and the evolution of the HI mass function and cosmic \HI\ density. In this paper, we present the science goals, the observing technique, the calibration and data reduction methods as applied to its initial data. In summary:
    
    \begin{enumerate}
    
        \item In order to quickly calibrate all 19 beams, we employ an on-the-fly (OTF) observation mode to scan a continuum calibrator source, which allows us to measure the flux density of the calibration noise diode, the system equivalent flux density, the pointing accuracy and the beam size.
        
        \item The high-temperature (10 K) noise diode was found to be stable, with a mean flux density of $0.72\pm0.08$ Jy. The mean system equivalent flux density was found to be $1.24\pm0.08$ Jy. The median and mean value of pointing errors were measured to be 6.3 and 8.1 arcsec, which is smaller than $\frac{1}{20}$ of FAST beam size at 1.5 GHz.
        
        \item The wide bandwidth of the multibeam receiver necessitated calculating the frequency dependence of calibration parameters in a manner that was robust against the considerable GNSS RFI present from 1.15 to 1.30 GHz. The main departures from frequency independence of calibration parameters were measured to be: a decrease in the high-frequency response of the noise diode; a low-frequency rise for the system equivalent flux density, and the expected inverse-frequency dependence for the beam size.
        
        \item Internal RFI was found to be problematic with early data. We were able to suppress this considerably by subtracting median spectra calculated in 19 beams over short periods of time. Further RFI flagging was developed to remove data contaminated by RFI, and ensure that strong \HI\ sources were not affected.
 
        \item For our initial target field data in FUDS0 and FUDS1, we employ the similar algorithms for bandpass calibration and imaging as for HIPASS and AUDS. The final cube for each FUDS field is 1 deg$\times$1 deg, with pixel size of 1 arcmin$\times$1 arcmin and frequency resolution of 22.9 kHz. Example spectra are presented, and a comparison with data from the Arecibo Ultra-Deep survey (AUDS) is made.
    \end{enumerate}

\section*{Acknowledgements}

    We would like to thank Jiguang Lu for his suggestions on calibration of 19 beams. Shijie Huang's kindly provided an RFI list for the FAST site. Pei Zuo helped improve the paper. 
    
    This work made use of the data from FAST (Five-hundred-meter Aperture Spherical radio Telescope). FAST is a Chinese national mega-science facility, operated by the National Astronomical Observatories, Chinese Academy of Sciences. The work is supported by the National Key R\&D Program of China under grant number 2018YFA0404703, and the Open Project Program of the CAS Key Laboratory of FAST, NAOC, Chinese Academy of Sciences. Parts of this research were supported by the Australian Research Council Centre of Excellence for All Sky Astrophysics in 3 Dimensions (ASTRO 3D), through project number CE170100013. BP would like to acknowledge the CAS-MPG LEGACY funding ``Low-Frequency Gravitational Wave Astronomy and Gravitational Physics in Space''.



\bibliography{References}

\appendix

\section{Beam model}

    \begin{table*}
        \caption{Parameterisation of the FAST multibeam beam sizes using Equation \ref{Equ_06} for both polarisations (XX$^*$ and YY$^*$) and their mean. Errors are given in parentheses.}
        \label{Tab_06}
        \centering
        \begin{tabular}{crrrrrrrrr}
            \hline\hline
            \multirow{3}{*}{Index}  & \multicolumn{9}{c}{Parameters}\\
            \cmidrule(lr){2-10}
            & \multicolumn{3}{c}{XX$^*$} & \multicolumn{3}{c}{YY$^*$} & \multicolumn{3}{c}{Mean}\\
            \cmidrule(lr){2-4} \cmidrule(lr){5-7} \cmidrule(lr){8-10}
            
            &         A [m] &         B [m] &           C [m]  &         A [m] &         B [m] &           C [m] &         A [m] &         B [m] &           C [m]\\
            \hline
            01 & 297.76 (0.01) & 61.90 (0.14) & -40.78 (0.38) & 301.86 (0.01) & 78.81 (0.16) & -74.51 (0.42) & 299.82 (0.01) & 71.30 (0.15) & -60.38 (0.40) \\
            02 & 288.59 (0.01) & 94.45 (0.23) & -61.65 (0.60) & 293.60 (0.01) & 93.13 (0.22) & -83.39 (0.59) & 291.02 (0.01) & 93.50 (0.23) & -71.19 (0.60) \\
            03 & 291.40 (0.01) & 76.97 (0.18) & -37.90 (0.47) & 296.05 (0.01) & 92.89 (0.19) & -85.81 (0.52) & 293.70 (0.01) & 84.99 (0.18) & -62.07 (0.49) \\
            04 & 293.63 (0.01) & 61.24 (0.19) & -19.72 (0.51) & 298.21 (0.01) & 73.76 (0.21) & -63.23 (0.58) & 295.92 (0.01) & 70.38 (0.20) & -49.37 (0.55) \\
            05 & 295.88 (0.01) & 81.11 (0.18) & -51.87 (0.48) & 300.22 (0.01) & 82.21 (0.18) & -78.12 (0.47) & 297.94 (0.01) & 81.89 (0.18) & -64.70 (0.48) \\
            06 & 303.30 (0.01) & 77.14 (0.25) & -63.12 (0.65) & 304.29 (0.01) & 67.62 (0.17) & -62.39 (0.47) & 303.83 (0.01) & 71.76 (0.21) & -61.44 (0.56) \\
            07 & 296.63 (0.01) & 72.65 (0.18) & -48.05 (0.47) & 302.12 (0.01) & 83.72 (0.19) & -84.37 (0.51) & 299.38 (0.01) & 77.72 (0.18) & -65.16 (0.49) \\
            08 & 277.80 (0.02) & 151.38 (0.38) & -109.88 (0.99) & 276.49 (0.02) & 107.02 (0.40) & -108.39 (1.04) & 277.15 (0.02) & 128.78 (0.39) & -109.83 (1.01) \\
            09 & 287.72 (0.02) & 143.71 (0.33) & -159.12 (0.84) & 289.49 (0.02) & 80.30 (0.32) & -76.65 (0.83) & 288.61 (0.02) & 111.98 (0.32) & -118.55 (0.84) \\
            10 & 291.24 (0.01) & 105.87 (0.24) & -84.27 (0.62) & 293.61 (0.01) & 70.85 (0.23) & -73.04 (0.59) & 292.42 (0.01) & 88.25 (0.23) & -78.90 (0.61) \\
            11 & 293.83 (0.01) & 62.22 (0.12) & -35.05 (0.34) & 298.91 (0.01) & 75.01 (0.13) & -64.02 (0.36) & 296.37 (0.01) & 67.90 (0.13) & -47.86 (0.35) \\
            12 & 297.79 (0.01) & 97.53 (0.23) & -71.84 (0.59) & 300.03 (0.01) & 71.05 (0.23) & -87.54 (0.60) & 298.91 (0.01) & 84.69 (0.23) & -81.30 (0.59) \\
            13 & 291.20 (0.02) & 140.28 (0.33) & -164.66 (0.85) & 292.74 (0.02) & 81.45 (0.34) & -87.82 (0.87) & 291.97 (0.02) & 110.76 (0.33) & -126.51 (0.86) \\
            14 & 289.79 (0.03) & 137.75 (0.48) & -119.92 (1.25) & 290.88 (0.03) & 75.58 (0.53) & -61.99 (1.40) & 290.28 (0.03) & 104.88 (0.51) & -86.65 (1.36) \\
            15 & 295.03 (0.02) & 115.27 (0.31) & -134.72 (0.79) & 296.77 (0.02) & 62.90 (0.30) & -66.03 (0.79) & 295.92 (0.02) & 88.47 (0.30) & -99.21 (0.79) \\
            16 & 302.00 (0.01) & 78.51 (0.23) & -57.53 (0.61) & 303.83 (0.01) & 45.72 (0.24) & -47.34 (0.61) & 302.92 (0.01) & 62.23 (0.23) & -53.19 (0.61) \\
            17 & 307.93 (0.01) & 56.37 (0.14) & -32.23 (0.37) & 314.11 (0.01) & 74.87 (0.16) & -87.36 (0.42) & 311.03 (0.01) & 65.28 (0.15) & -59.20 (0.39) \\
            18 & 301.42 (0.01) & 88.80 (0.27) & -71.80 (0.70) & 305.13 (0.02) & 57.18 (0.28) & -54.07 (0.72) & 303.27 (0.02) & 73.57 (0.28) & -64.53 (0.71) \\
            19 & 291.82 (0.02) & 123.85 (0.32) & -142.91 (0.81) & 294.22 (0.02) & 70.24 (0.31) & -83.66 (0.78) & 293.03 (0.02) & 96.48 (0.31) & -113.35 (0.80) \\
            \hline\hline
        \end{tabular}
    \end{table*}
    

\label{lastpage}

\end{document}